\documentclass[useAMS,usenatbib,referee]{mn2e}
\usepackage[dvipdfmx]{graphicx}
\usepackage{pdfpages}
\usepackage{amsmath}  % Advanced maths commands
\usepackage{amssymb}  % Extra maths symbols
\usepackage{multirow}
\newcommand{\etal }{{et al.} }

\newcommand{\msun}{{\rm M}_\odot}

\newcommand{\cc}{{\rm cm^{-3}}}

\title
[Failed and Delayed Protostellar Outflows]%[**/45 characters]
{Failed and delayed protostellar outflows with high mass accretion rates}

\author[Machida \& Hosokawa]{Masahiro N. Machida$^{1,2}$ and Takashi Hosokawa$^{3}$
\\
Department of Earth and Planetary Sciences, Faculty of Sciences, Kyushu University, Fukuoka, Fukuoka 819-0395, Japan\\
Department of Physics and Astronomy, The University of Western Ontario, London, ON N6A 3K7, Canada\\
Department of Physics, Graduate School of Science, Kyoto University, Sakyo-ku, Kyoto 606-8502, Japan
}
\date{Accepted XXX. Received YYY; in original form ZZZ}
\pubyear{2019}

\begin{document}
\label{firstpage}
\pagerange{\pageref{firstpage}--\pageref{lastpage}}
\maketitle

\begin{abstract}%[***/200 words for Letters]
The evolution of protostellar outflows is investigated under different mass accretion rates in the range $\sim10^{-5}$--$10^{-2}\,\msun$\,yr$^{-1}$ with three-dimensional magnetohydrodynamic simulations. 
 A powerful outflow always appears in strongly magnetized clouds with $B_0 \gtrsim B_{\rm 0, cr}$ $=10^{-4}  (M_{\rm cl}/100\,\msun)$\,G, where $M_{\rm cl}$ is the cloud mass. 
When a cloud has a weaker magnetic field, the outflow does not evolve promptly with a high mass accretion rate. 
In some cases with moderate magnetic fields $B_0$ slightly smaller than $B_{\rm 0,cr}$, the outflow growth is suppressed or delayed until the infalling envelope dissipates and the ram pressure around the protostellar system is significantly reduced. 
In such an environment, the outflow begins to grow and reaches a large distance only during the late accretion phase. 
On the other hand, the protostellar outflow fails to evolve and is finally collapsed by the strong ram pressure when a massive ($\gtrsim 100\msun$) initial cloud is weakly magnetized with $B_0 \lesssim 100$\,$\mu$G. 
The failed outflow creates a toroidal structure that is supported by magnetic pressure and encloses the protostar and disk system. 
Our results indicate that high-mass stars form only in strongly magnetized clouds, if all high-mass protostars possess a clear outflow.
If we would observe either very weak or no outflow around evolved protostars, it means that strong magnetic fields are not necessarily required for high-mass star formation.
In any case, we can constrain the high-mass star formation process from observations of outflows.
\end{abstract}
%%%%%%%%%%%%%%%%%%%%%%%%%%%%%%%%%%%%%%%%
\section{Introduction}
\label{sec:intro}
Protostellar outflows can provide information on the very early phase of star formation. 
A protostar is born as a consequence of the gravitational collapse of a molecular cloud core. 
The protostellar mass immediately after formation is $\sim10^{-3}$--$10^{-2}\,\msun$ \citep{larson69,masunaga00}. 
Then, the protostar grows by mass accretion from the infalling envelope that  is the remnant of the molecular cloud core. 
The mass accretion (or the released gravitational energy of the accreting matter) drives the protostellar outflow. 
Thus, a protostellar outflow is considered to be promising evidence for the accretion phase of star formation. 

High mass accretion rates are expected in high-mass star formation, and they result in driving powerful and massive outflows.
Massive outflows with large linear momenta and kinetic energies are usually observed around high-mass protostars \citep[e.g.,][]{beuther02,wu04,zhang05,maud15,villiers14,villiers15,li18}. 
\citet{wu04} showed a clear correlation between outflow kinetic energies and bolometric luminosities of star-forming cores \citep[see also][]{beuther02,zhang05,maud15,li18}.
If the bolometric luminosity is dominated by the accretion luminosity $L_{\rm acc} \propto M_* \dot{M}$, it can be used as an index to evaluate the protostellar mass $M_*$ and accretion rate $\dot{M}$. In observations, both the outflow kinetic energy and linear momentum increase as the bolometric luminosity increases.

In the framework of low-mass star formation, \citet{tomisaka98} showed by magnetohydrodynamical (MHD) simulations that the protostellar outflow is magnetically driven \citep[see also][]{tomisaka00,tomisaka02}. 
The magnetically driven outflow in the high-mass star formation process has also been investigated in several studies \citep[e.g.,][]{banerjee07,commercon11, hennebelle11, seifried12,tanaka17,tanaka18,kolligan18}. 
Our previous simulation studies, \citet[][hereafter Paper I]{matsushita17} and \citet[][hereafter Paper II]{matsushita18}, showed that the mass ejection rate is proportional to the mass accretion rate, as suggested by observations.
A massive outflow tends to appear in the high-mass star-formation case 
because the released gravitational energy is large when the mass accretion rate is high.
Note that the gravitational energy is converted to the outflow kinetic energy through a magnetic effect \citep{blandford82,tomisaka02}, and a high-mass star is formed with a high-mass accretion rate \citep[e.g.][]{beuther07}.   
However, in previous studies, the outflow driving  was investigated in a limited parameter range, in which the magnetic energy of the prestellar cloud is assumed to be comparable to the gravitational energy.

Observations have shown that in nearby star forming regions the magnetic energy is comparable to the gravitational energy \citep[e.g.,][]{crutcher99,troland08}.
The mass-to-flux ratio normalized by its critical value is usually used as an index of the degree of magnetization for a prestellar cloud, which is defined as 
\begin{equation}
\mu_0 = \left( \frac{M_{\rm cl}}{\Phi_{\rm cl}} \right) / \left( \frac{1}{2\pi G^{1/2}} \right),
\label{eq:mu}
\end{equation}
where $M_{\rm cl}$ and $\Phi_{\rm cl}$ are the mass and magnetic flux of the prestellar cloud, respectively. 
\citet{troland08} showed that the mass-to-flux ratio is $\mu_0 \sim2$--$3$ on average, which means that the magnetic field cannot halt a prompt gravitational collapse of the prestellar core.
When the magnetic field is as strong as $\sim 10 \mu$\,G, the magnetic energy is almost of the same order as the gravitational energy of a typical prestellar cloud core for low-mass star formation. 
However,  the mass and gravitational energy are large in prestellar cores for high-mass star formation \citep{zhang09,pillai11,tan13,sanhueza17}.
Thus,  the magnetic energy becomes relatively low compared with the gravitational energy when  high-mass star-forming clouds are of the same order as the magnetic field strength (e.g. $\sim10\mu$\,G, for details, see \S\ref{sec:settings}). 

In simulation studies,  very strong magnetic fields or small mass-to-flux ratios  have been conventionally assumed for prestellar clouds for high-mass star formation.
However, a magnetic field strength of $\sim100$--$1000\mu$\,G is necessary to obtain $\mu_0 \sim2$ for a high-mass prestellar cloud core, as described in \S\ref{sec:settings}.
It is very difficult to estimate the magnetic field strength of  high-mass prestellar clouds, because high-mass star-forming regions are located far from the sun.
If the magnetic field strength in a high-mass prestellar core is $\sim 10~\mu {\rm G}$, as in the low-mass cases, the mass-to-flux ratio becomes $\mu_0 \gg10$ and the magnetic energy is over $100$--$1000$ times smaller than the gravitational energy. 
Such a relatively low magnetic energy would change the picture of star formation, because the magnetic field plays several pivotal roles in the star formation process, relating to angular momentum transfer, collapse geometry, formation of rotationally supported and pseudo disks, and the driving of protostellar outflows \citep{inutsuka12}. 
We briefly explained that no outflow appears in weakly magnetized clouds in Papers I and II, where we mainly focused on cases with strongly magnetized clouds.

The protostellar outflow has a significant impact on the star formation in general \citep[e.g.,][]{matzner00}. 
In this study, following on from Papers I and II, we investigate high-mass star formation in rotating magnetized cloud cores from a broader point of view. To comprehensively understand the role of the magnetic field and outflow driving, we further study the evolution in both strongly and very weakly magnetized clouds with a wide range of parameters. 

The rest of the paper is structured as follows. \S2 describes the numerical settings of our model, and the simulation results are presented in \S3. We discuss the outflow driving in weakly magnetized clouds and compare our results with observations in \S4. A summary is presented in \S5.

%%%%%%%%%%%%%
%%% Table1%%%
%%%%%%%%%%%%%
\renewcommand{\arraystretch}{1.2}
\begin{table}
\begin{center}
\begin{tabular}{c||cccccccccc} \hline
Model & $f$ &  $\mu_0$ & $B_0$ [G] & $\Omega_0$ [s$^{-1}$] & $\alpha_0$ & $\beta_0$ & $\gamma_0$  & $M_{\rm cl}$ \, [$\msun$]  & Outflow & frag.\\
\hline
AM2 & 1.4 & 2  & $1.5\times10^{-5}$ & \multirow{5}{*}{$3.2\times10^{-14}$}  & \multirow{5}{*}{0.5} & \multirow{5}{*}{0.02} & 0.19 &\multirow{5}{*}{11} & Successful & N \\
AM3 &  1.4 & 3  &$1.0\times10^{-5}$  & & & & 0.083 & &  Successful & N\\
AM5 &  1.4 & 5  &$6.2\times10^{-6}$ & & & & 0.030 & &  Delayed & N\\
AM10 & 1.4 & 10&$3.1\times10^{-6}$ & & & & $7.5\times10^{-3}$  & & Delayed & N\\
AM20 & 1.4 & 20&$1.6\times10^{-6}$ & & & & $1.9\times10^{-3}$  & &  Delayed & N\\
\hline
BM2 & 3.4 & 2  & $3.8\times10^{-5}$ & \multirow{5}{*}{$5.0\times10^{-14}$}  & \multirow{5}{*}{0.2} &\multirow{5}{*}{ 0.02}& 0.19 &\multirow{5}{*}{28} & Successful & N\\
BM3 & 3.4 & 3&$2.5\times10^{-5}$ & & & & 0.083 & &   Successful & N\\
BM5 & 3.4 & 5&$1.5\times10^{-5}$ & & & &0.030 & &  Delayed & N\\
BM10 & 3.4 & 10&$7.5\times10^{-6}$ & & & & $7.5\times10^{-3}$ & &  Failed  & N\\
BM20 & 3.4 &20 &$3.7\times10^{-6}$ & & & & $1.9\times10^{-3}$ & &   Failed & N\\
\hline
CM2 & 8.4 & 2  & $9.2\times10^{-5}$ & \multirow{5}{*}{$7.9\times10^{-14}$}  & \multirow{5}{*}{0.08} & \multirow{5}{*}{0.02}& 0.19  &\multirow{5}{*}{68} & Successful & N\\
CM3 & 8.4 & 3& $6.2\times10^{-5}$ & & & & 0.083 & &   Successful & N\\
CM5 & 8.4 & 5& $3.7\times10^{-5}$ & & & & 0.030 & &    Failed & N\\
CM10 & 8.4 & 10& $1.9\times10^{-5}$ & & & & $7.5\times10^{-3}$ & &  Failed & N\\
CM20 & 8.4 &20 &$6.2\times10^{-6}$ & & & &$1.9\times10^{-3}$ & &  Failed & N\\
\hline
DM2 & 16.8 & 2  & $1.9\times10^{-4}$ & \multirow{5}{*}{$1.1\times10^{-13}$}  & \multirow{5}{*}{0.04} & \multirow{5}{*}{0.02} &  0.19 &\multirow{5}{*}{132} & Successful & N\\
DM3 & 16.8 & 3& $1.2\times10^{-4}$  & & & & 0.083 & &  Successful & N\\
DM5 & 16.8 & 5& $7.4\times10^{-5}$ & & & & 0.030 & & Failed & N\\
DM10 & 16.8 & 10&$3.7\times10^{-5}$ & & & &  $7.5\times10^{-3}$ & & Failed & Y (2)\\
DM20 & 16.8 &20 &$1.8\times10^{-5}$ & & & &$1.9\times10^{-3}$ & & Failed & N\\
\hline
EM2 & 33.6 & 2  & $3.7\times10^{-4}$ & \multirow{5}{*}{$1.5\times10^{-13}$}  & \multirow{5}{*}{0.02} & \multirow{5}{*}{0.02} & 0.19 &\multirow{5}{*}{272}  & Successful & N \\
EM3 & 33.6 & 3&$2.4\times10^{-4}$ & & & & 0.083 & &  Successful & N\\
EM5 & 33.6 & 5&$1.5\times10^{-4}$ & & & & 0.030 & &  Delayed & Y (4) \\
EM10 & 33.6 & 10&$7.4\times10^{-5}$ & & & & $7.5\times10^{-3}$  & & Failed & Y (4) \\
EM20 & 33.6 &20 &$3.7\times10^{-5}$ & & & &$1.9\times10^{-3}$ & &  Failed & Y (4) \\
\hline
FM2 & 67.2 & 2  & $7.4\times10^{-4}$ & \multirow{5}{*}{$2.2\times10^{-13}$}  & \multirow{5}{*}{0.01} & \multirow{5}{*}{0.02} & 0.19 &\multirow{5}{*}{545}& Successful & Y (2) \\
FM3 & 67.2 & 3&$5.0\times10^{-4}$ & & & & 0.083 & &   Delayed & Y (6) \\
FM5 & 67.2 & 5&$3.0\times10^{-4}$ & & & & 0.030 & &  Delayed & Y (6)\\
FM10 & 67.2 & 10& $1.5\times10^{-4}$ & & & & $7.5\times10^{-3}$ & &  Delayed & Y (4)\\
FM20 & 67.2 &20 & $7.4\times10^{-5}$ & & & &$1.9\times10^{-3}$ & &  Failed & Y (8) \\
\hline
\end{tabular}
\end{center}
\caption{
Model name, initial cloud parameters and calculation results.
Column 1 gives the model name. 
Columns 2 and 3 give the parameters $f$ and $\mu_0$. 
Columns 4 and 5 give the magnetic field strength $B_0$ and angular velocity $\Omega_0$ for the initial state. 
Columns 6--8 give the ratios of the thermal $\alpha_0$, rotational $\beta_0$ and magnetic $\gamma_0$ energies to the gravitational energy of the initial cloud. 
Column 9 gives the initial cloud mass.
Column 10 describes the calculation results, in which `Successful', `Delayed' and `Failed' mean that an outflow successfully appears, a delayed outflow appears and an outflow fails to appear, respectively. 
Column 11 describes whether fragmentation occurs (Y) or not (N), in which the total number of fragments is described in parenthesis.
}
\label{table:1}
\end{table}

\section{Initial Conditions and Numerical Settings}
\label{sec:settings}
The initial conditions and numerical settings are almost the same as in Papers I and II,
and thus, we only briefly explain them in this section.
As the initial state, we adopt a spherical cloud (core) with a Bonnor--Ebert  (B.E.) density profile with a central density of  $n_{\rm c,0}=10^5\, \cc$ and an isothermal temperature of $T_{\rm iso,0}=20$\,K.
The initial cloud  has twice the critical B.E. radius ($R_{\rm cl}=4.1\times10^4$\,au).
To promote gravitational collapse, the cloud density is enhanced by $f$, which is the density enhancement factor (Papers I and II) and is related to the cloud stability $\alpha_0$ (the ratio of the thermal to gravitational energy).
Thus, the initial cloud has a central density of $n_{\rm cl}= f \times 10^5\,\cc$ (=$f \times n_{\rm c,0}$). 
A uniform density $n_{\rm ISM}=n_{\rm cl}/80.0$ is set outside the initial cloud ($r>R_{\rm cl}$) to mimic the interstellar medium.  

A rigid rotation $\Omega_0$ and uniform magnetic field $B_0$ are added to the initial cloud, with the magnetic field direction set to be parallel to the rotation axis or the $z$-axis. 
As described in Table~\ref{table:1}, 30 different prestellar clouds are prepared as the initial state, in which the density enhancement factor $f$ (or $\alpha_0$) and the mass-to-flux ratio $\mu_0$ (or magnetic field strength $B_0$) are parameters, where $f$ controls the mass accretion rate (see, \S\ref{sec:results} and Papers I and II) and $\mu_0$ determines the degree of magnetization. 
We adopt  $f=1.4$, 3.4, 8.4, 16.8, 33.6 and 67.2, and the cloud mass and cloud stability $\alpha_0$  differ accordingly among the models.
As described in Table~\ref{table:1}, the initial cloud has a mass in the range of $M_{\rm cl} = 11$--$545\,\msun$.

The initial magnetic field strength $B_0$ is adjusted so that the resulting mass-to-flux ratio is $\mu_0=2$, 3,  5, 10 and 20 in each cloud. 
Since $\mu_0$ depends also on the cloud mass, the magnetic field strength for the initial cloud $B_0$ differs depending on both $f$ and $\mu_0$. 
The rotation rate $\Omega_0$, which is also listed in Table~\ref{table:1}, is determined to give $\beta_0=0.02$ in each model.

The model name and the ratio of the magnetic to gravitational energy $\gamma_0$ are also listed in Table~\ref{table:1}. 
The parameter $f$ adopted in this study is exactly the  same as that in Papers I and II.
The difference between this study and Papers I and II is the magnetic field strength (or the mass-to-flux ratio). 
$\mu_0=2$ is adopted in almost all models in Papers I and II,  while five different ratios  $\mu_0=2$, 3, 5, 10, 20 are adopted in this study. 
Note that the cloud parameters are almost the same between this study and Papers I and II except for the magnetic field strengths, while we changed the initial isothermal temperature from $T_{\rm iso}=40$\,K (Paper I and II) to $T_{\rm iso}=20$\,K (this study) when constructing the B.E. density profile. 
Thus, the physical quantities for the initial clouds differ somewhat between this study and our previous studies.  

The numerical settings are also the same as in Papers I and II. 
We solve the resistive MHD equations including  self-gravity (eqs.~[1]--[4] of \citealt{machida19} and eq.~[2] of Paper I), in which the barotropic equation of state (eq.~[1] of Paper I) is  used.
The nested grid code is used to cover a wide density and spatial range \citep[for details of the code, see][]{machida04,machida05a,machida10,machida13}. 
At the beginning of the calculation, five levels of nested grids are prepared. 
Each grid is composed of (i, j, k) = (64, 64, 32), and mirror symmetry is imposed on the  $z=0$ plane.

The initial prestellar cloud is embedded in the fifth level of the grid ($l=5$). 
The first level of the grid has a box size of $L(l=1)=6.6\times10^5$\,au and a cell width of $h(l=1)=1.0\times10^4$\,au.  
Both the box size and cell width halve with each increment of grid level $l$. 
The computational boundary  corresponding to the surface of the $l=1$ grid is located $2^4$ times further from the surface of the initial cloud, which  can suppress artificial reflection of the Aflv\'en wave at the boundary \citep[for details, see][]{machida13}.
A new finer grid is generated to satisfy the Truelove condition, in which the Jeans length is resolved with at least 16 cells \citep{truelove97}. 
The maximum grid level is set to $l_{\rm max}=15$, which has a box size of $L(l=l_{\rm max})= 40$\,au and cell width of $h(l=l_{\rm max})=0.62$\,au.

The sink method is used to calculate the main accretion phase for a long duration, in which a threshold density of $ n_{\rm thr}=10^{13} \cc$ and sink radius $r_{\rm sink}=1$\,au are adopted.
The sink parameters are  the same as those in Papers I and II. 
With these settings, we calculate the time evolution for $10^4$\,yr after the protostar formation (or sink creation) for all the models.

\section{Results}
\label{sec:results}

\subsection{Mass Accretion Histories}
The differences between low- and high-mass star formation in the early evolutionary phase are characterized by  the mass accretion rate.
The mass accretion rate is expected to be higher in  high-mass star formation than in  low-mass star formation. 
As described in \S\ref{sec:settings},  in this study, we changed the cloud stability $\alpha_0$ using $f$ to control the mass accretion rate. 
The mass accretion rate onto a protostar is roughly proportional to $\propto \alpha_0^{-3/2}$ (for details, see Paper I), which indicates that a model with a larger $f$ (or smaller $\alpha_0$)  has a higher mass accretion rate. 

Figure~\ref{fig:1} plots the mass accretion rate against the protostellar mass for each model. 
The mass accretion rate has a peak around $M_{\rm ps}\sim 0.2$--$0.3\,\msun$ for models with $f=1.4$, 3.4, 8.2 and 16.8. 
The initial enhancement in the mass accretion rate is attributed to the existence  of the first core formed prior to the protostar formation \citep{larson69,masunaga00}. 
Since  the first core (remnant) remains around the protostar even shortly after protostar formation, the protostar acquires its mass from the first core during the very early mass accretion phase \citep{machida11}. 
The temperature in the first core ($\sim100$\,K) is higher  than that in the infalling envelope ($\sim10$\,K) and the mass accretion rate is roughly proportional to $\propto  c_s^3$ when the host object (or cloud) for  the protostar  is in a nearly equilibrium state \citep{larson03}, where $c_s$ is the sound speed. 
Thus, the mass accretion rate is high while the first core remnant remains around a protostar. 
Then, the mass accretion decreases after the first core remnant disappears because the protostar acquires its mass  directly from the infalling envelope that has a low temperature of $\sim10$\,K.
For the models with large $f$ (or smaller $\alpha_0$), the protostar forms without the formation of a long-lived first core (Paper I,  \citealt{bhandare18}). 
Thus, we do not observe a significant enhancement in the mass accretion rate during the early accretion phase in such models (models with $f=33.6$ and 67.2).

Figure~\ref{fig:1} indicates that there is no significant difference in the mass accretion rate among the models with the same $f$ and $\alpha_0$, although the oscillation of the mass accretion rates can be confirmed in each panel. 
To easily confirm the history of the mass accretion rate, the protostellar mass for each model is plotted against the elapsed time after protostar formation in Figure~\ref{fig:2}. 
The protostellar mass differs during the very early phase ($t_{\rm ps}\lesssim 10^2$--$10^3$\,yr) among the models  with the same parameter $f$.
However, for $t_{\rm ps} \gtrsim10^3$\,yr, the models with the same $f$ have almost the same protostellar mass. 
The difference in the protostellar mass among the models with the same $f$ is within a factor of two at $t_{\rm ps}\simeq10^4$\,yr. 
Thus, the initial difference in magnetic field strength does not significantly change the mass accretion rate.  
On the other hand, both the mass accretion rate (Fig.~\ref{fig:1}) and protostellar mass (Fig.~\ref{fig:2}) differ considerably among  the models with different $f$ (or $\alpha_0$). 

To emphasize the difference in the mass accretion rate among the models with different $f$,  the accretion rates with different $f$ but the same mass-to-flux ratio $\mu_0=2$ (upper) and $10$ (lower) are plotted against the protostellar mass (left) and the elapsed  time after protostar formation (right) in Figure~\ref{fig:3}. 
We can see a clear difference  in the mass accretion rate among the models with different $f$ in Figure~\ref{fig:3}. 
The mass accretion rate at $t_{\rm ps}\sim10^4$\,yr for models with $f=1.4$ (models AM2 and AM10) is $\dot{M}\sim3\times10^{-5}\,\msun$\,yr$^{-1}$, while for  models with $f=67.2$ it is $\dot{M}\sim10^{-2}\,\msun$\,yr$^{-1}$. 
Thus, the difference in the mass accretion rate between models with $f=1.4$ and 67.2 is over two orders of magnitude. 
Almost the same difference and trend were also confirmed in Paper I (see Fig.~2 of Paper I). 
Thus, we can investigate different cloud evolutions from the viewpoint of the mass accretion rate with models with  different $f$ (or $\alpha_0$). 

\subsection{Typical Models}
In this subsection, we focus on the models with the same $f$ (or $\alpha_0$),  but with different $\mu_0$ (or $B_0$). 
Figure~\ref{fig:4} shows the cloud evolution for models DM2 (left column), DM5 (middle column) and DM10 (right column), which have the same parameters $f=16.8$ and $\alpha_0=0.04$.
The initial magnetic field strengths of these models are $B_0 = 1.9\times10^{-4}$ (DM2), $7.4\times10^{-5}$ (DM5) and $3.7\times10^{-5}$\,$\mu$G (DM10). 
Figure~\ref{fig:4} left column shows that the outflow gradually evolves with time and has a size of $\sim10^4$\,au at $t_{\rm ps}\simeq10^4$\,yr (Fig.~\ref{fig:4}{\it d}). 
The outflow has a well collimated structure at each time (Figs.~\ref{fig:4}{\it a}--{\it d}). 
In addition, we can see that a large fraction of the cloud gas is ejected by the outflow in Figure~\ref{fig:4}{\it d}.
Thus, for model DM2, the outflow would have a significant impact on the star formation. 

For model DM5 (Fig.~\ref{fig:4} middle column), the outflow has a size of $\sim300$\,au at $t_{\rm ps}=545.2$\,yr (Fig.~\ref{fig:4}{\it e}), then it shrinks with time (Fig.~\ref{fig:4}{\it f}--{\it h}). 
We manage to confirm a bipolar structure in the center of the cloud for model DM 5 at $t_{\rm ps}\simeq10^4$\,yr (Fig.~\ref{fig:4}{\it h}).
However, we cannot confirm a noticeable outflow in the right column of Figure~\ref{fig:4}, which describes the  cloud evolution  for model DM10.
As described above, the difference in the magnetic field strength between models DM2 ($1.9\times10^{-4}$\,$\mu$G) and DM10 ($3.7\times10^{-5}$\,$\mu$G) is about a factor of five. 
Thus, Figure~\ref{fig:4} indicates that a slight difference in the magnetic field strength of prestellar clouds dramatically changes the outflow driving and star formation process.

To further investigate the models that do not show a mature outflow (models DM5 and DM10), the cloud evolutions at a small scale for these models are plotted in Figures~\ref{fig:5} and \ref{fig:6}. 
Figure~\ref{fig:5} shows the time sequence of the outflow for model DM5. 
For this model, the outflow gradually evolves for $t_{\rm ps}\lesssim 3000$\,yr (Figs.~\ref{fig:5}{\it a} and {\it b}).
Then, the outflow stagnates with  a size of $\sim1000$\,au. 
The outflow widens or expands in the horizontal direction with time and a cavity-like structure appears (Figs.~\ref{fig:5}{\it e}-{\it f}) during $t_{\rm ps}\gtrsim3000$\,yr. 
However, the outflow is not very active and does not grow in the vertical direction  by the end of the simulation. 

Figure~\ref{fig:6} shows the time sequence of the outflow for model DM10, in which only the density and velocity distributions in the region around the protostar within $\lesssim300$\,au are plotted. 
For this model, although a bipolar outflow-like structure develops, it does not extend to  a large distance. 
Instead, the bipolar structure, which has a size of $200$--$300$\,au, oscillates and collapses, spreading out in the horizontal direction (Figs.~\ref{fig:6}{\it e}--{\it g}).
The outflow (or bubble) like structure seen in Figure~\ref{fig:6} does not disappear,  while  the oscillation of the outflow recurrently occurs in the horizontal and vertical directions by the end of the simulation.

\subsection{Outflow Evolution and Classification}
In this subsection, we investigate  the time evolution of  outflows for  all the models and categorize them into three types.  
Figure~\ref{fig:7} shows the outflow size $L_{\rm out}$ for all models against the elapsed time after protostar formation. We identified the outflow as the region where the radial velocity $ v_r$ exceeds $v_{\rm cri}$ and $v_{\rm cri}  = 1$\,km\,s$^{-1}$ is adopted.  
Note that although  we adopted  $v_{\rm cri}  = 0.1$, 0.5 and 3\,km\,s$^{-1}$ to confirm the dependence on $v_{\rm cri}$, the outflow physical quantities did not significantly depend on  $v_{\rm cri}$ in this range.  
The outflow size is determined as being the farthest outflowing region from the protostar (or the center of the cloud).

Figure~\ref{fig:7} shows that the outflow size is larger in models with smaller $\mu_0$ than in models with larger $\mu_0$.
In other words, the outflow size  is larger in strongly magnetized clouds  than in weakly magnetized clouds.
Thus, outflows appeared in  strongly magnetized clouds and reached a large distance in a short time. 

As seen in each panel of Figure~\ref{fig:7}, by the end of the simulation ($t_{\rm ps}\simeq10^4$\,yr),  the outflows in models with  $\mu_0=2$ and 3 reach $L_{\rm out} \sim10^4$\,au, which is comparable to the initial cloud radius $R_{\rm cl}=4.1\times10^4$\,au. 
We call models showing an outflow that continues to grow by the end of the simulation   `Successful outflows.'
Model DM2 shown in the left column of Figure~\ref{fig:4} is a typical example of a successful outflow.
We categorized  models AM2, AM3,  BM2, BM3, CM2, CM3, DM2, DM3, EM2, EM3 and  FM2  as being successful outflows.

%%On the other hand,  in the model with $f=1.4$ and $\mu_0=10$ (AM10, Fig.~\ref{fig:7}{\it a}), the outflow  grows at $\sim2\times10^3$\,yr and  shrinks during $t_{\rm ps}\gtrsim 7\times10^3$\,yr.
Figure~\ref{fig:7}{\it b} shows that  the outflows do not grow significantly in the models with $\mu_0=10$ and 20 (BM10 and BM20). 
The oscillation of $L_{\rm out}$ seen in these models indicates that the outflow transiently shrinks after it slightly  grows. 
We call this type of outflow a `Failed outflow.' 
As seen in Figures~\ref{fig:5} and \ref{fig:6}, for the failed outflow case the outflow fails to grow and does not reach a large distance by the end of the simulation.
For these models, the outflow stays within a region $\lesssim 10^3$\,au from the protostar and does not show significant growth in the vertical direction during  $\sim10^4$\,yr after protostar formation.
Thus, the outflow lengths at $t_{\rm ps}=10^4$\,yr are much smaller than the initial cloud radius $R_{\rm cl}=4.1\times10^4$\,au, which means that the outflow is deeply embedded in the infalling envelope for a long time.  
We categorized models BM10, BM20, CM5, CM10, CM20, DM5, DM10, DM20, EM10, EM20 and FM20 as failed outflows.

In addition to successful and failed outflows, there exists another type of outflow. 
In models EM5 (Fig.~\ref{fig:7}{\it e}), FM3, FM5 and FM10 (Fig.~\ref{fig:7}{\it f}), the outflow does not grow during the early accretion phase ($t_{\rm ps}\lesssim 10^3$\,yr), while  it begins to grow significantly during the later accretion phase ($t_{\rm ps}\gtrsim  10^3$\,yr).
We call this type of outflow a `Delayed outflow.'
In the models with $f=33.6$ (Fig.~\ref{fig:7}{\it e}) and $67.2$ (Fig.~\ref{fig:7}{\it f}), when the magnetic field strength of the initial cloud is as weak  as $\mu_0=5$, 10 and 20, the outflow exponentially grows only in the later accretion phase. 
In these models, the outflow does not evolve in $t_{\rm ps}\lesssim10^3$\,yr, during which the outflow size is  within $\sim100$\,au. 
Then, the outflow suddenly grows and the outflow size begins to increase for $t \gtrsim  10^3$\,yr. 
Although the outflow size for these models is  $L_{\rm out}\sim 10^2$--$10^3$\,au at the end of the simulation, the outflow appears to evolve in a further evolutionary stage. 
Thus, the outflow evolution is `delayed' in these models. 
We confirmed that models AM5, AM10  AM20 and BM5 also show a similar evolution to models EM5, FM3, FM5 and FM10. 
We categorized models AM5, AM10,  AM20, BM5, EM5, FM3, FM5 and FM10 as delayed outflows.

To clearly see the evolution of delayed outflows, we show time sequences in Figures~\ref{fig:8} and \ref{fig:9}.
In model FM10 (Fig.~\ref{fig:8}), we cannot confirm a noticeable outflow for  $t_{\rm ps}\lesssim 9000$\,yr (Figs.~\ref{fig:8}{\it a}--{\it c}).
However, the outflow rapidly grows and has a size of $\sim4000$\,au (Figs.~\ref{fig:8}{\it e}--{\it f}) for  $t_{\rm ps}\gtrsim  9000$\,yr (Figs.~\ref{fig:8}{\it e}--{\it g}). 
The outflow evolution for model EM5 is plotted in Figure~\ref{fig:9}. 
For this model, after the outflow reaches $\sim500$\,au (Fig.~\ref{fig:9}{\it a}), it shrinks to a size of $\sim100$\,au (Fig.~\ref{fig:9}{\it b}). 
The outflow slowly grows and reaches $\sim10^3$\,au at $t_{\rm ps}\simeq 9000$\,yr (Fig.~\ref{fig:9}{\it c}). 
Then, the outflow exponentially grows in a further evolutional  stage (Figs.~\ref{fig:9}{\it e}--{\it g}). 
Figures~\ref{fig:8} and \ref{fig:9} also show that  the density of the infalling envelope that encloses the outflow gradually decreases. 
The density of outflow (cavity) is over one order of magnitude larger than that of the infalling envelope (Fig.~\ref{fig:8}{\it e}--{\it f} and Fig.~\ref{fig:9}{\it e}--{\it f}).
For a delayed outflow, the outflow does not grow while the infalling envelope is dense, and it then grows significantly as the infalling envelope dissipates. 
As described above, we classified the models into three categories: Successful, Failed and Delayed outflows. 
The classification for each model is described in Table~\ref{table:1}. 

\subsection{Outflow Physical Quantities}
The outflow physical quantities are plotted in Figures~\ref{fig:10}--\ref{fig:12}, in which the outflow mass (Fig.~\ref{fig:10}), momentum (Fig.~\ref{fig:11}), and momentum flux (Fig.~\ref{fig:12}) are plotted against the elapsed time after the protostar formation. 
Figure~\ref{fig:10} shows the outflow mass, which is calculated as
\begin{equation}
M_{\rm out} = \int^{v_r > v_{\rm cri}} \rho \, dV,
\end{equation}
plotted against the elapsed time after protostar formation for all models, in which the range of the outflow mass (i.e. the $y$-axis) is adjusted to emphasize the difference between the models with different $\mu_0$ in each panel. 
The figure indicates that the outflow mass is large when the initial magnetic field is strong. 
Independent of the parameter $f$ (or $\alpha$), the outflow mass in the model with $\mu_0=2$ and 3 is about one or two orders of magnitude larger than the models with $\mu_0=10$ and 20. 

We can see the same trend in the outflow momentum $P_{\rm out}$ (Fig.~\ref{fig:11}) and momentum flux $F_{\rm out}$ (Fig.~\ref{fig:12}) as in the outflow mass (Fig.~\ref{fig:12}).
They are estimated as 
\begin{equation}
P_{\rm out}   = \int^{v_r > v_{\rm cri}} \rho \, v_r \, dV,
\end{equation}
\begin{equation}
\label{eq:fout_theory}
F_{\rm out} = \dfrac{P_{\rm out}}{t_{\rm ps}}. 
\end{equation}
As seen in Figures~\ref{fig:7} and \ref{fig:10}, there is  a gap  in the outflow momentum and momentum flux between strongly and weakly magnetized clouds. 
In Figure~\ref{fig:11}, we can confirm that the outflow momentum in the models EM10, EM20 and FM10  rapidly increases for $t_{\rm ps}\gtrsim 10^3$\,yr. 
However, the outflow momenta in these models are still about one order of magnitude smaller than those in the strongly magnetized models (EM2, EM3, FM2 and FM3). 

In each panel of Figure~\ref{fig:12}, we can see a significant difference in the outflow momentum flux.
The outflow momentum flux has a dimension of force, and is used as an index of the outflow driving force. 
Figure~\ref{fig:12} indicates that the outflow momentum flux differs among the models with the same $f$ from the beginning (or the protostar formation epoch). 
Thus, the initial difference in the magnetic field strength causes the difference in the outflow driving force. 
Independent of $f$, the outflow momentum flux is largest for the models with $\mu_0=2$ and 3, and smallest for the models with  $\mu=10$ and $20$. 
The former corresponds to the successful outflow case, while the latter corresponds to the failed or delayed outflow case. 
The models with a moderate magnetic field strength $\mu_0=5$ show different behaviors, dependent on $f$. 

\subsection{Spiral Structure and Fragmentation Induced by Gravitational Instability}
\label{sec:spiral}
When the magnetic field of the prestellar  cloud is strong, the angular momentum around the protostar is efficiently transported by magnetic effects such as  magnetic braking and outflow \citep{zhao20}.
As a result, a small-sized (rotationally supported) disk appears  around the protostar \citep{machida19b}.
On the other hand, when the magnetic field is weak, the angular momentum is not significantly  transported by the magnetic effects.
In such a case, the infalling gas gradually accumulates in the circumstellar region  and a large-sized circumstellar disk appears  \citep{machida11b}.
After the circumstellar disk becomes massive, gravitational instability occurs \citep{toomre64,machida10}, leading to the formation of a spiral structure that may eventually fragment 	(see Paper I and \S\ref{sec:disk}).

Figure~\ref{fig:13} shows the time sequence of the circumstellar disk  for models DM2, DM5 and DM10, which are the same as in Figure~\ref{fig:4}.
For these models,  a spiral structure develops without  fragmentation in models DM2 and DM5, while fragmentation occurs and two clumps appear around the cloud center in model DM10. 
The figure also indicates that the disk size increases as the initial magnetic field weakens, in which  a prominent spiral or fragment appears in models with relatively weak magnetic fields.
A comparison of Figure~\ref{fig:4} and Figure~\ref{fig:13} indicates  that  the outflow activity seem to be anti-correlated with the disk size.
It is expected that the angular momentum around the protostar is not effectively transported by the outflow  (and magnetic braking) in weakly magnetized clouds because an excess angular momentum promotes the disk growth.

Both the spiral structure and the orbital motion of fragments create an anisotropic gravitational field.
Thus, in these models,  the angular momentum transport should also be due to the gravitational torque, in addition to the magnetic effects such as outflow, magnetic braking and magneto-rotational instability \citep[for details, see][]{machida19b}. 
As a result, the mass accretion rates onto the protostar for the models with the same $f$ (or $\alpha_0$) but with different $\mu_0$ (or $B_0$) are almost the same (Fig.\ref{fig:1}), because the gravitational torque also plays an important role for the angular momentum transport  when the magnetic field is weak.  
Therefore,  the difference in magnetic field strength does not significantly affect the determination of the mass accretion rate and the protostellar growth.
 Since the disk gravitational instability in both strongly and weakly magnetized clouds with  high mass accretion rates  was already  investigated in Papers I and II, we do not particularly focus on it in this study. 
We simply describe the disk properties in \S\ref{sec:disk}.

\section{Discussion}
\subsection{Suppression of Outflow with High Ram Pressure}
\label{sec:ramP}
As described  in \S\ref{sec:results},  the outflow growth is suppressed during the main accretion phase in weakly magnetized clouds. 
However, in some models, the outflow begins to grow in the later main accretion phase. 
Figures~\ref{fig:6} and \ref{fig:8} show that the outflow develops only after the density of the infalling envelope becomes significantly low. 
Thus, it is expected that the outflow growth is suppressed due to the existence of a (dense) infalling envelope. 

Figure~\ref{fig:14} plots the configuration of magnetic field lines within the outflow region for model DM10 at the same epoch as in Figure~\ref{fig:6}{\it e}. 
After this epoch, the outflow begins to shrink as shown in Figures~\ref{fig:6}{\it f} and {\it g}. 
Figure~\ref{fig:14} shows  that the magnetic field lines are strongly twisted around the $z$-axis inside the outflow. 
Thus, the gas inside the outflow region should be pushed out in the vertical direction (or positive $z$ direction) due to the strong magnetic pressure gradient force if  the external pressure outside the outflow can be ignored. 
The outflow region, however, is actually pushed into the central region (or negative $z$ direction) as time proceeds (Figs.~\ref{fig:6}{\it f} and {\it g}). 
The mass accretion rate is  high in the models with a large $f$ or small $\alpha_0$.
Thus, it is expected  that  the outflow driving is suppressed by the strong ram pressure of the infalling envelope.

To quantitatively investigate  the suppression of the outflow, we compare the ram pressure with the magnetic pressure for models DM2, DM5 and DM10. 
Each panel of Figure~\ref{fig:15} plots the ram pressure outside the outflow and the magnetic pressure inside the outflow, where both the ram  $P_{\rm ram}= \rho v^2 /2$ and magnetic $P_{\rm mag}=B^2/8 \pi$ pressures are normalized by the thermal pressure at the center of the initial cloud $P_{\rm th, 0} $.  
We confirmed that  the thermal pressure can be ignored compared with the ram and magnetic pressures in the region around the outflow.

Figures~\ref{fig:15}{\it a}--{\it c} indicate that for model DM2 (left column), the magnetic pressure inside the outflow is considerably larger than the ram pressure outside the outflow at any epoch. 
Therefore, the outflow could continue to grow until the end of the simulation for this model. 
On the other hand, the ram pressure outside the outflow is comparable or larger than the magnetic pressure inside the outflow for models DM5 (Fig.~\ref{fig:15} middle column) and DM10 (Fig.~\ref{fig:15} right column).  
It is expected that the ram pressure suppresses the outflow growth in these models. 
For example, we focus on model DM10.
For this model, Figure~\ref{fig:15}{\it h} corresponds to the epoch shown in Figure~\ref{fig:14}. 
At this epoch, the magnetic field lines are strongly twisted (Fig.~\ref{fig:14}), while the magnetic pressure inside the outflow is much smaller than the ram pressure outside the outflow. 
As a result, the outflow in model DM10 does not appreciably evolve by the end of the simulation. 
However, the outflow does not completely disappear, because the magnetic field can be amplified as the outflow shrinks.
For this model, the outflow continues to oscillate like a spring with a size of $<1000$\,au during the main accretion phase.

During the main accretion phase, the magnetic field is mainly amplified by the rotation of the circumstellar disk, as shown in Figure~\ref{fig:14}, while it dissipates due to Ohmic dissipation in the disk. 
Since the formation and evolution of the circumstellar disk are closely related to the strength of the magnetic field \citep{zhao20}, it is not easy to clarify the evolution of the magnetic field and the outflow driving condition.
However, Figure~\ref{fig:15} indicates that the high ram pressure interrupts the outflow growth for the models with initially weak magnetic fields. 
It is natural that the outflow begins  to grow after the infalling envelope dissipates (Fig.~\ref{fig:9})  because the ram pressure is proportional to the density,  which decreases with time in the infalling envelope \citep{larson03}. 
Thus, we can conclude that a high ram pressure, which is realized with a high-mass accretion rate (see Paper I), suppresses the outflow growth. 

When the magnetic field of the initial cloud is as strong as $\mu_0\lesssim5$, the outflow promptly grows without being disturbed by the ram pressure during the main accretion phase. 
The gas of the infalling envelope is dense and the ram pressure is high on a small scale ($\sim100-1000$\,au), while the ram pressure decreases as the distance from the center of the cloud increases. 
Thus, once the outflow extends to a large distance cutting through the ram pressure barrier, the outflow is not suppressed by the ram pressure. 
On the other hand, when the initial cloud has a weak magnetic field $\mu_0 \gtrsim 5$,  the outflow will not grow until the infalling envelope dissipates. 
In this case, the outflow growth is significantly delayed. 
Alternatively, the outflow fails  to grow by the end of the mass accretion phase when the initial magnetic field is as weak as $\mu_0\gtrsim10$. 
Note that since the outflow is powered by the mass accretion, the outflow loses its activity as the infalling envelope dissipates \citep{machida13}. 
Although we have followed a long-term evolution for $\sim 10^4$\,yr, the outflow might nevertheless begin to grow in a further late stage even in the failed cases. 

\subsection{Comparison with Observations}
\label{sec:obs}
In this subsection, we compare our simulation results with observations. 
We have already performed such comparisons in Papers I and II, in which we considered the simulation results only for the cases with $\mu_0=2$. 
In Papers I and II, we showed  that time derivative quantities such as the outflow momentum flux $F_{\rm out}$ and mechanical  luminosity $L_{\rm kin}$ are useful for the comparison, because  there is still a gap in the evolutionary stage (or  the outflow dynamical timescale) between our simulations and observations. 
We calculated the outflow evolution for $\sim10^4$\,yr in this study as in  Papers I and II, although a part of the massive outflows observed in high-mass star forming regions has a longer dynamical timescale exceeding $10^4$\,yr. 
We can then compare the time derivative outflow quantities, which are essential to understand the nature of the outflows \citep{bontemps96}, between the simulations and observations as described in Paper II. 
The outflow momentum fluxes and mechanical luminosities obtained from both the simulations and observations are plotted against the stellar bolometric luminosity in Figure~\ref{fig:16}. 
\citet{wu04} derived the  fitting formula of the outflow momentum flux  $F_{\rm out}$, for many samples of observed outflows \citep[e.g.,][]{,beuther02, zhang14, maud15},
\begin{equation}
{\rm log}\, (F_{\rm out}/(M_{\rm sun} \, {\rm km}\, {\rm s}^{-1}\, {\rm yr}^{-1})  )  = -4.92 +  0.648\, {\rm log}\,(L_{\rm bol}/L_{\rm sun}), 
\label{eq:fout}
\end{equation} 
and the outflow mechanical  luminosity $L_{\rm kin}$, 
\begin{equation}
{\rm log}\, (L_{\rm kin}/L_{\rm sun}) = -1.98 + 0.62\, {\rm log}\,(L_{\rm bol}/L_{\rm sun}). 
\label{eq:eout}
\end{equation} 
These  are also plotted by the solid line in each panel of Figure~\ref{fig:16}.
The colored symbols represent the evolution in the MHD simulations, for which the stellar bolometric luminosity was calculated by numerically solving the protostellar evolution with the STELLAR code \citep{yorke08,hosokawa13}.
In our current treatment, we followed the protostellar evolution over the variable accretion histories taken from the simulations as post-processes \citep[see also Paper II and][]{machida13}. 
The evolution of the stellar radius and luminosity is described in appendix \S\ref{pevol}.
The detailed settings and calculation methods are given in Paper II.
The derivation of the outflow momentum flux in the simulations is described in equation~(\ref{eq:fout_theory}). 
To derive the outflow mechanical luminosity, we estimated the outflow kinetic energy,
\begin{equation}
E_{\rm out} = \int^{v_r > v_{\rm cri}} \rho\, v^2 dV,
\end{equation}
where $v=(v_x^2 + v_y^2 + v_y^2)^{1/2}$, and the outflow mechanical luminosity is estimated as 
\begin{equation}
L_{\rm kin}= \frac{E_{\rm out}}{t_{\rm ps}},
\end{equation} 
where $t_{\rm ps}$ is the elapsed time after protostar formation. 
The derivation of the outflow momentum flux and kinetic energy are the same as in Paper II.

The left and middle panels of Figures~\ref{fig:16} indicate that both the outflow momentum flux and mechanical luminosity estimated  from the simulation roughly agree with the observations when the initial clouds have  $\mu_0=2$ and 5. 
The outflow momentum fluxes for the models with $\mu_0=2$ are distributed above the solid line  (eq.~\ref{eq:fout}) in the range $L_{\rm bol}\lesssim 10^3\,L_{\rm sun}$, while they are slightly below the solid line  in the range $L_{\rm bol}>10^5L_{\rm sun}$.
Figure~\ref{fig:16}{\it a} shows that although the simulation data  slightly deviate from equation~(\ref{eq:fout}), they are within the range of observations of \citet{beuther02},  \citet{zhang14} and  \citet{maud15}.
The outflow momentum fluxes for the models with $\mu_0=5$ (Fig.~\ref{fig:16}{\it b}) show a similar distribution to the models with $\mu_0=2$.
The trend of the outflow mechanical luminosity for the models with $\mu_0=2$ and 5 (Fig.~\ref{fig:16}{\it d} and {\it e}) is almost the same as that of the outflow momentum flux (Figs.~\ref{fig:16}{\it a} and {\it b}). 

On the other hand, both the outflow momentum flux and mechanical luminosity for the model with $\mu_0=10$ are considerably smaller than the observations. 
In Figures~\ref{fig:16}{\it c} and {\it f}, although only three points taken from the simulation are distributed above the solid line (eq.~\ref{eq:eout}), almost all the data points taken from the simulations are below the line.
In addition, in Figure~\ref{fig:16}{\it c}, the simulation results are significantly smaller than the observations of \citet{beuther02},  \citet{zhang14} and  \citet{maud15}.
Thus, the simulated outflows in weakly magnetized clouds are considerably weaker than the observations. 

\subsection{Outflow Driving Condition}
\label{sec:condition}
A clear outflow does not always appear in the high-mass star formation process, because a high ram pressure due to the high mass accretion rate suppresses the outflow growth when the magnetic field  of the prestellar cloud is not strong.
In this subsection, we quantitatively discuss the necessary condition for outflow driving.

As shown in \S\ref{sec:results}, independent of the initial cloud mass $M_{\rm cl}$, a powerful outflow  appears immediately after or before protostar formation  when the initial magnetic field is as strong as $\mu_0 \le 3$ (successful outflow models, see Table~\ref{table:1}).  
We can relate the initial magnetic field  $B_0$  to the cloud mass $M_{\rm cl}$ with equation~(\ref{eq:mu}), in which the cloud radius  $R_{\rm cl}=4.1\times10^4$\,au is introduced  to derive the initial magnetic field  $B_0 = \Phi_{\rm cl}/\pi R_{\rm cl}^2$. 
Using the initial normalized mass-to-flux ratio $\mu_0$ (\S\ref{sec:intro}), the relation between $B_0$ and  $M_{\rm cl}$ can be described as 
\begin{equation}
B_0 = \dfrac{3\times10^{-4}}{\mu_0} \left( \dfrac{M_{\rm cl}}{100\msun} \right) \, {\rm G }. 
\end{equation}
Thus, for the simulation results with  $\mu_0 \lesssim 3$, the outflow driving condition can be represented as $ B_0 \gtrsim  B_{\rm cr, 0} $ where 
\begin{equation}
 B_{\rm cr, 0} = 10^{-4}   \left( \frac{M_{\rm cl}}{100\msun} \right) \, {\rm G }. 
\label{eq:condition}
\end{equation}
A massive outflow appears if this condition is satisfied.
The condition also indicates that the magnetic field  necessary for the outflow driving  is proportional to the initial cloud mass.
For example, a magnetic field of $B_0 \gtrsim 10^{-4}\,{\rm G}$ is necessary for outflow driving in a cloud with $M_{\rm cl}=100\,\msun$. 
It should be noted that, in our settings, the cloud mass is related to the mass accretion rate by the parameter $\alpha_0$, and a high mass accretion rate is realized in an initially high mass cloud (for details, see Paper I).  

Although  condition (\ref{eq:condition}) was derived  with `successful' outflow models, the outflow appears also  in `delayed' outflow models. 
The mass-to-flux ratio or magnetic field strength necessary for the delayed outflow differs among the models with different initial cloud masses (see Table~\ref{table:1}).
Thus, we cannot clearly describe an outflow driving condition that includes the delayed outflow models. 
However, we can estimate the initial magnetic field strength from observations with outflow physical quantities and bolometric luminosity.
The outflow momentum flux and kinetic luminosity in the successful outflow models are roughly fitted with equations~(\ref{eq:fout}) and (\ref{eq:eout}), while  those in the delayed outflow models are smaller than the successful models, as seen in Figure~\ref{fig:16}. 
The difference in the outflow physical quantities between the successful and delayed models can be attributed to the initial magnetic field strength (see Table~\ref{table:1}). 
Thus, based on the successful models and  condition~(\ref{eq:condition}), we can roughly estimate the magnetic field strength of the initial cloud.
If the observed outflow quantities are distributed around the fitting line in Figure~\ref{fig:16},  condition (\ref{eq:condition}) should be fulfilled for the initial conditions for high-mass star formation.  
On the other hand, if we can confirm  the outflows distributed below the successful models or fitting lines of Figure~\ref{fig:16} in future observations, this would indicate that the magnetic field of the prestellar cloud should be smaller than $ B_0 <10^{-4}(M_{\rm cl}/100\, \msun)\,{\rm G}$. 
Observations to date indicate that the high-mass stars form only in strongly magnetized clouds (see \S\ref{sec:scenario}).

\subsection{High-mass Star Formation  and Outflow Driving}
\label{sec:scenario}
Understanding the high-mass star formation process is a  hotly debated topic in the star formation field.  
Recent studies have suggested that high-mass stars form by the same framework as low-mass star formation.
Some researchers consider that the high-mass star formation is a scaled-up version of low-mass formation, which  is called the  core accretion scenario \citep{tan14}. 
In this scenario,  the high-mass star forms in a gravitationally collapsing cloud core, and the circumstellar disk and outflow appear during the main accretion phase \citep{tan16}. 
The massive outflows observed in high-mass star forming regions are  considered to be definitive evidence of core accretion  in the high-mass star formation process, because the massive outflow is proof of the occurrence of active mass accretion \citep{beuther02}.

However, in this study, we showed that even when a high-mass protostar forms according to the core accretion scenario, a massive outflow does not always appear. 
Thus, it is not clear whether a massive outflow is a good indicator to identify  a high-mass star formation site.  
As described in \S\ref{sec:results}, during the early main accretion phase, a massive outflow appears only when the host cloud for the high-mass star formation is very strongly magnetized. 
When the magnetic field of the star-forming cloud is not very strong,  the outflow is weak and would not be observable. 
In this sense, the massive outflow may trace a part of the whole high-mass star formation.

As described in \S\ref{sec:obs}, the observations of massive outflows agree well with our simulation results for strong magnetic fields ($\mu_0=2$ and 5). 
We note that for strong magnetic fields there are no data in the lower right domain where the bolometric luminosity is large but both the outflow momentum flux and mechanical luminosity are small (see Fig.~\ref{fig:16}).
On the other hand, when a high-mass star forming cloud is weakly magnetized, the simulation data are distributed there (Figs.~\ref{fig:16}{\it c} and {\it f}). 
We can observe compact outflows using large telescopes such as ALMA, 
and if it is the case that no data is observed in this region, it is considered that high-mass star formation occurs only in clouds with very strong magnetic fields.

\citet{inoue13} pointed out that high-mass star formation occurs only in a shocked compressed layer where the magnetic field is greatly enhanced, in which a strong shock is caused by a cloud--cloud collision or a collision between flows in a highly turbulent environment.
They also showed that the gravitational collapse does not begin until a sufficient amount of gas flows into the shocked region and the self-gravity dominates  the Lorentz-force.  
In this scenario, the gravitational collapse begins just after the gravitational energy, which is proportional to the cloud mass, is comparable to the magnetic energy  \citep[see also][]{vaidya13,inoue18}.
Thus, the magnetic field  must be  strong.
Since high-mass prestellar cloud cores are strongly magnetized, massive outflows should always appear in this scenario. 

If weak outflows are observed around high-mass protostars in future observations, this would indicate that massive outflows were  preferentially observed in past observations. 
This would suggest that there is a wide variety of magnetic field strengths in high-mass star formation.  
Some recent observations have reported very weak outflows around evolved protostars in low-mass star forming regions \citep{tokuda18,aso19}. These might be examples of low-mass star formation associated with very weak magnetic fields, possible low-mass counterparts of high-mass failed or delayed cases. It is clear that there is more to discover about star formation from high-resolution observations of protostellar outflows. 

\subsection{Effect of Sink Cell and High-velocity Jet}
As described in \S\ref{sec:settings}, to realize a long-term evolution, we introduced a sink cell that covers the region around the protostar.
Thus, we cannot resolve the protostar and inner disk region in our calculations. On the other hand, it is considered that high-velocity jets appear near the protostar by the star-disk magnetosphere interaction \citep[e.g.,][]{shu94,matt05,matt08,machida06,arce07}.
However, recent observations indicate that the wide-angle low-velocity outflow is driven by the disk outer region much far from the protostar and the contribution of the high-velocity jets is limited \citep{bjerkeli16,alves17,zhang18,matsushita19}.
\citet{hirota17} observed the rotation of the protostellar outflow in a high-mass star-forming region and found the outflow 
launching region far from the protostar, while they could not find any signatures of the high-velocity jet launched from the disk inner part.

Since we cannot observe the jet driving region with a sufficiently high spatial resolution, the jet  driving mechanism has not been fully established. 
However, many clear jets were observed in star forming regions \citep[e.g.][]{ray20} and they would affect the low-velocity outflow.
In this study, we investigated  the failed and delayed outflow cases, in which the outflow activity is weak.
The high-velocity  jets may activate such a weak outflow.
Although we cannot calculate a long-term evolution without sink, we need to investigate the effect of the high-velocity jets in future studies. 

\section{Summary}
In this study, we investigated the evolution of protostellar outflows in high-mass star formation, performing a suite of resistive MHD simulations. 
We studied the dependence of the evolution on the mass accretion rates and magnetic field strength, particularly considering cases with weakly magnetized clouds. 
%%--------------------------------------------------
We showed that a massive outflow, which are frequently observed in high-mass star-forming regions, does not always appear when the mass accretion rate is high. 
A massive outflow appears only when the prestellar cloud is strongly magnetized and the magnetic energy  is equivalent to the gravitational energy in the prestellar cloud  (the `successful outflow' case).  
The properties of simulated outflows emerging from such clouds agree well with observations. An outflow in a successful case reaches $\sim10^4$\,au after the evolution for $\sim10^4$\,yr after the protostar formation.

On the other hand, when the magnetic field of the prestellar cloud is weak, the outflow growth is suppressed by the high ram pressure of the rapid accretion flow in the envelope. 
We classified the outflow evolution in weakly magnetized clouds into two categories, failed and delayed outflows.
In both the failed and delayed outflow cases, the outflow does not significantly evolve for $\sim10^4$\,yr after the birth of the protostar.
Although the outflow reaches $\sim100$--$1000$\,au from the protostar, it  stagnates in both cases. 
In the delayed outflow case, the outflow can ultimately grow, overcoming the ram pressure barrier. 
The ram pressure cannot suppress the outflow once the outflow sufficiently evolves, because the ram pressure decreases 
as the outflow extends to the outer part of the infalling envelope.  
On the other hand, in the failed outflow cases, the outflow does not grow and creates a torus-like structure with a size of $\sim1000$\,au around the protostellar system of the protostar and circumstellar disk.

Our results indicate that a subtle difference in the initial magnetic field strength of the high-mass star forming cloud can cause a significant difference in the outflow growth. 
The magnetic field strength differs by less than a factor of 5 between successful ($\mu_0=2$, 3, 5) and delayed/failed ($\mu_0\ge5$) outflow cases. Since observations have identified many massive outflows \citep[e.g.,][]{beuther02}, it has been considered that massive outflows universally appear in high-mass star formation. If this is correct, then high-mass star formation necessarily occurs only in strongly magnetized clouds where the field strength is $\mu_0=2$--$5$.
If this is not the case, then we are missing weak outflows driven from the high-mass protostars owing to current observational limitations. Since high-mass star-forming regions are located far from the sun, it is difficult to observe small-sized weak outflows.
In any case, the possible diversity of protostellar outflows is a key to understanding high-mass star formation. We expect that future high-spatial resolution observations of outflows will further reveal the high-mass star formation process.

\section*{Acknowledgements}
This research used the computational resources of the HPCI system provided by the Cyber Science Center at Tohoku University, the Cybermedia Center at Osaka University, and the Earth Simulator at JAMSTEC through the HPCI System Research Project (Project ID: hp180001, hp190035, hp200004).
Simulations reported in this paper were also performed by 2019 and 2020 Koubo Kadai on the Earth Simulator (NEC SX-ACE) at JAMSTEC. 
The present study was supported  by JSPS KAKENHI Grants (JP17H02869, JP17H06360, JP17K05387, JP17KK0096: MM, JP17H01102, JP19H01934: TH).

\section*{DATA AVAILABILITY}
The data underlying this article are available in the article and in its
online supplementary material.

\clearpage
%%%%%%
% Fig. 1
%%%%%%
\begin{figure*}
\begin{center}
\includegraphics[width=1.0\columnwidth]{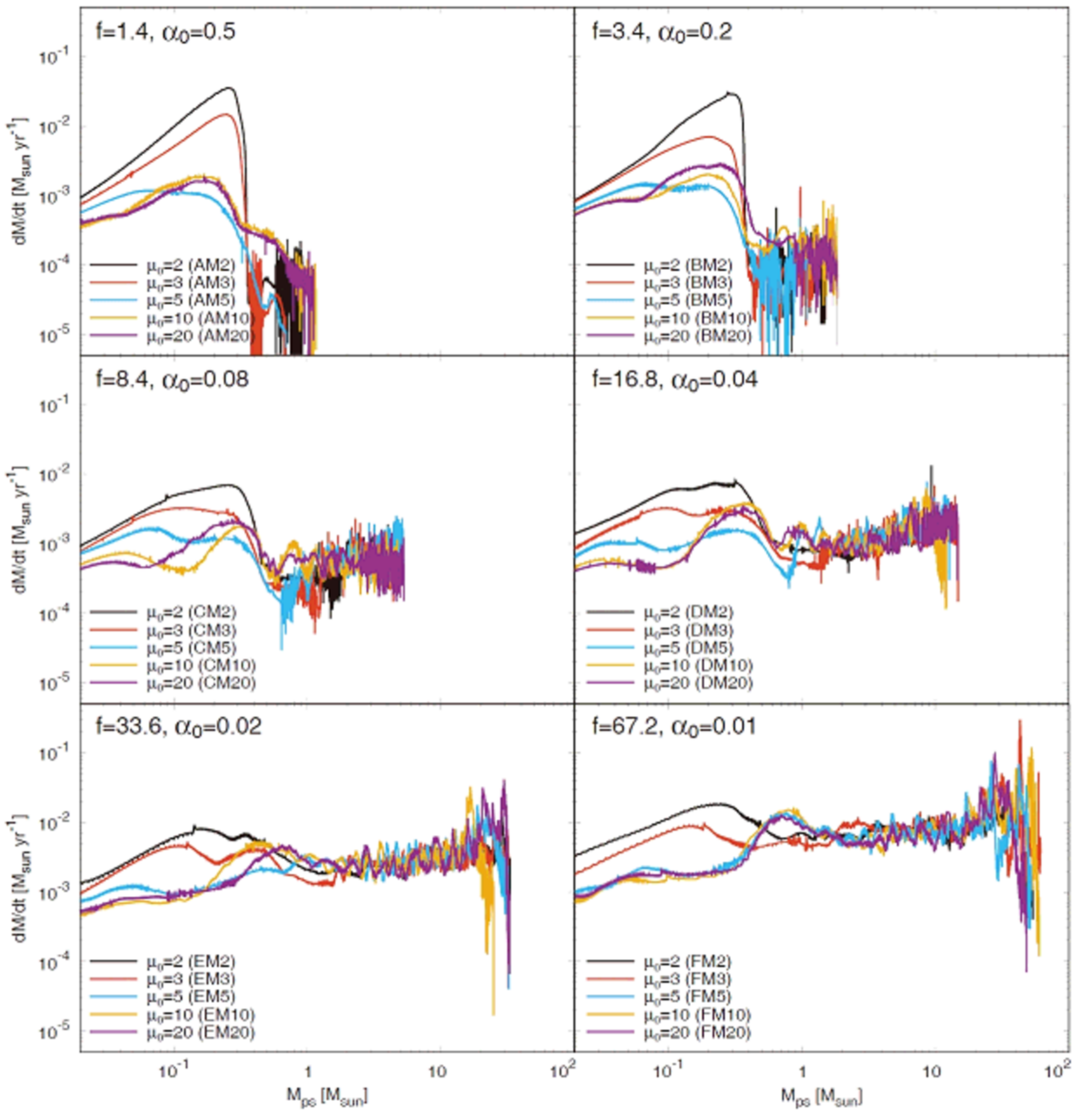}
\end{center}
\caption{
Mass accretion rate for all models against protostellar mass $10^4$\,yr after protostar formation. 
The density enhancement factor $f$ (or $\alpha_0$) is the same for all the models in each panel, while the mass-to-flux ratio $\mu_0$ differs. 
The initial $\mu_0$, model name and parameters $f$ and $\alpha_0$ are described in each panel.
}
\label{fig:1}
\end{figure*}

%%%%%%
% Fig. 2
%%%%%%
\begin{figure*}
\begin{center}
\includegraphics[width=1.0\columnwidth]{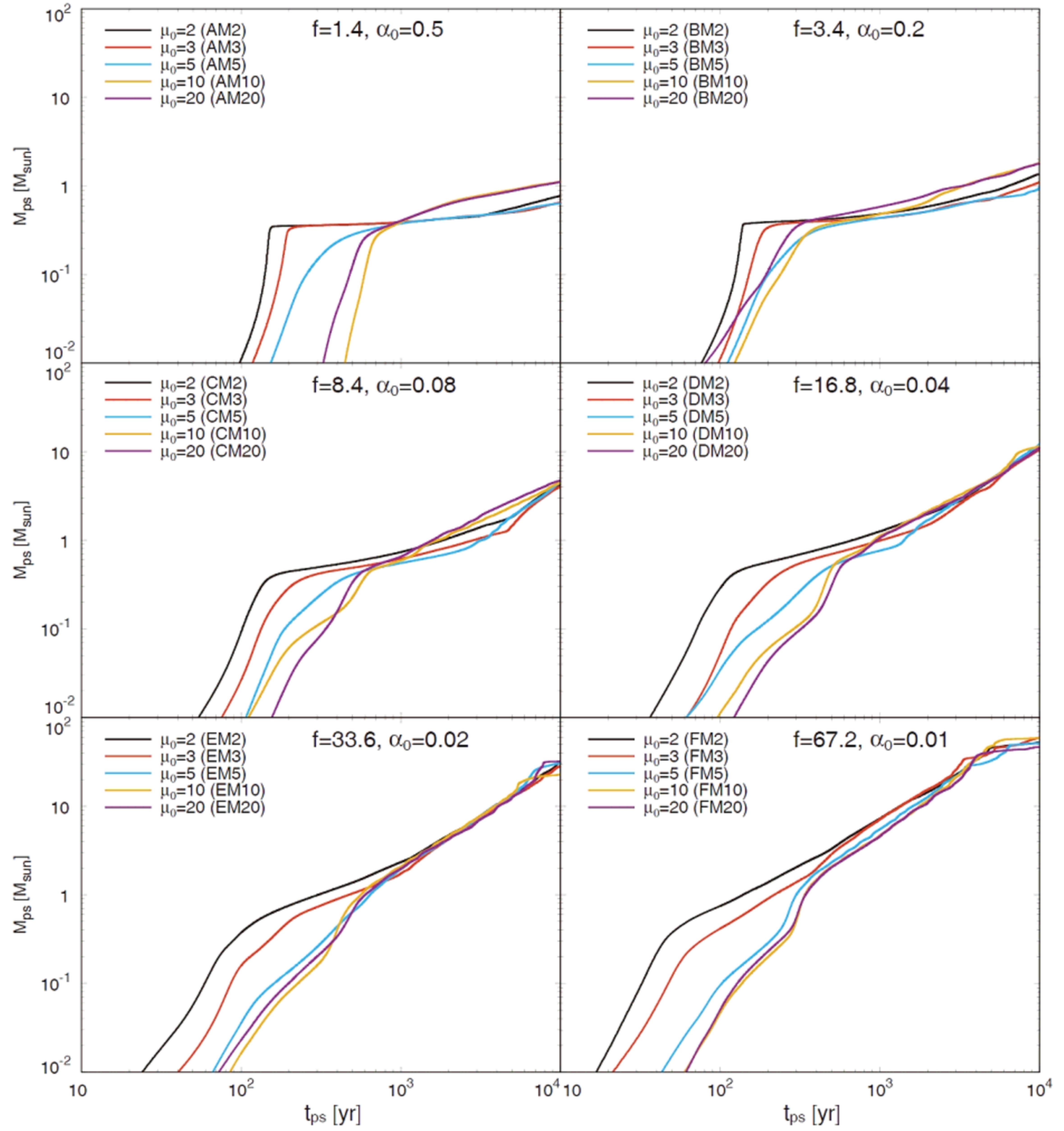}
\end{center}
\caption{
Protostellar mass for all models against the elapsed time after protostar formation. 
The model name and parameters $f$ and $\alpha_0$ are given for each model. 
}
\label{fig:2}
\end{figure*}

%%%%%%
% Fig. 3
%%%%%%
\begin{figure*}
\begin{center}
\includegraphics[width=1.0\columnwidth]{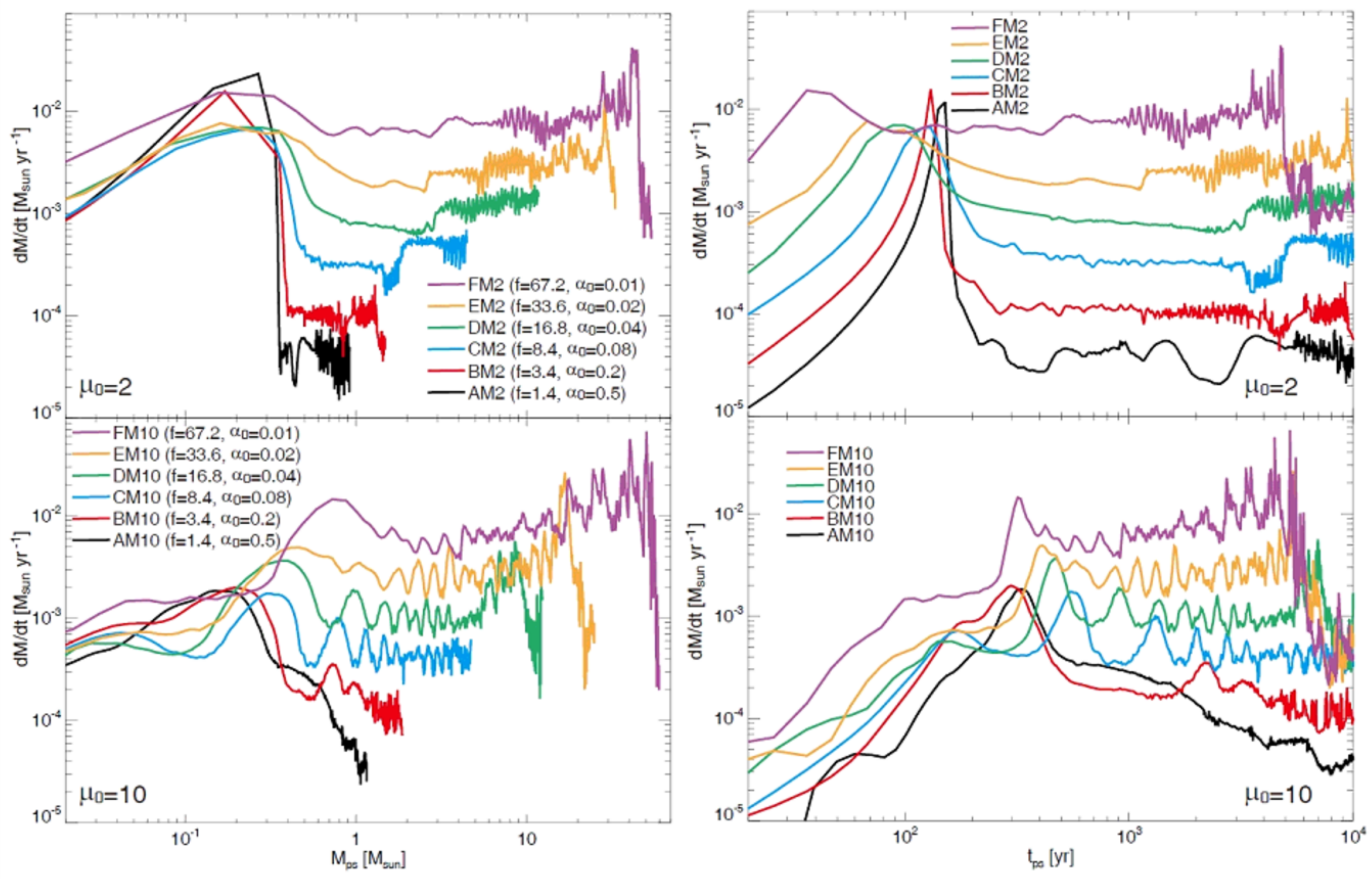}
\end{center}
\caption{
Mass accretion rate for models AM2, BM2, CM2, DM2, EM2, FM2 (top), AM10, BM10, CM10, DM10, EM10 and FM10 (bottom) against the protostellar mass (left) and the elapsed time after protostar formation (right). 
}
\label{fig:3}
\end{figure*}

%%%%%%
% Fig. 4
%%%%%%
\begin{figure*}
\begin{center}
\includegraphics[width=0.9\columnwidth]{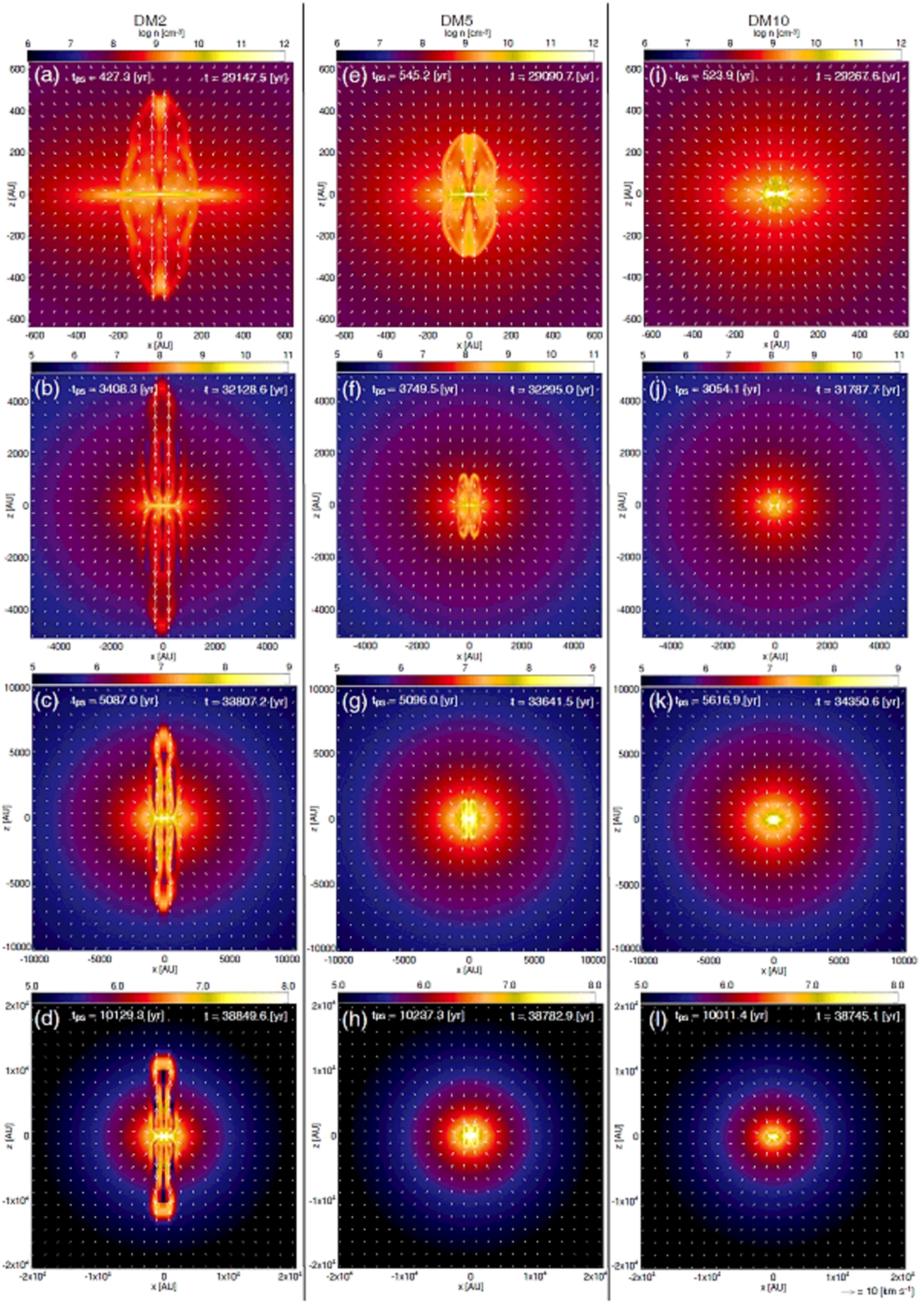}
\end{center}
\caption{
Time sequence of density (color) and velocity (arrows) distributions on the $y=0$ plane for models DM2 (left; $f=16.8$ and $\mu_0=2$, see also movie DM2.avi), DM5 (middle; $f=16.8$ and $\mu_0=5$) and DM10 (right; $f=16.8$ and $\mu_0=10$). 
The elapsed time after protostar formation $t_{\rm ps}$ and that after the cloud begins to collapse $t$ are described in each panel. 
The spatial scale differs in each row. 
}
\label{fig:4}
\end{figure*}

%%%%%%
% Fig. 5
%%%%%%
\begin{figure*}
\begin{center}
\includegraphics[width=1.0\columnwidth]{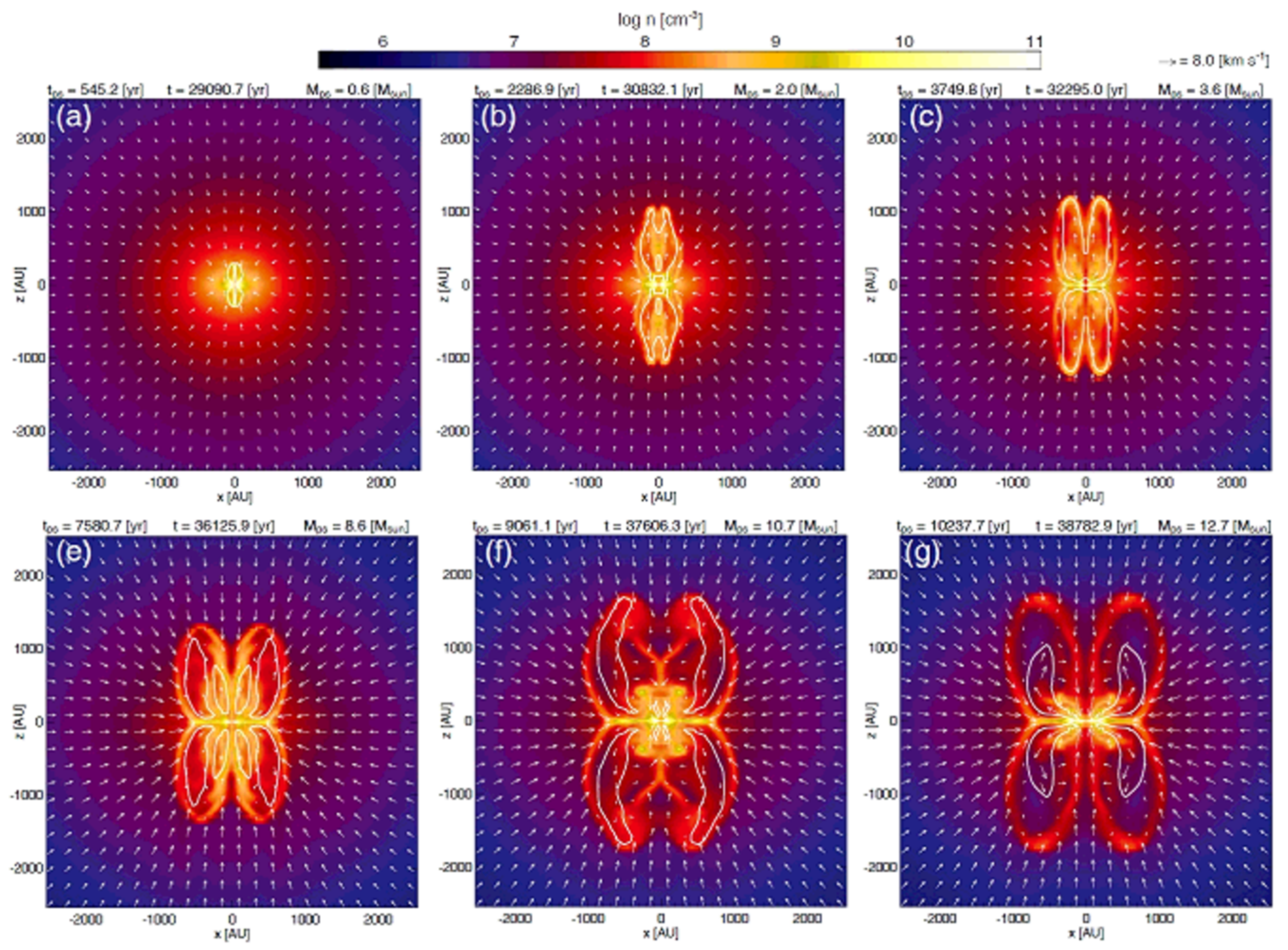}
\end{center}
\caption{
Time sequence of model DM5 ($f=16.8$ and $\mu_0=5$), classified as a `failed outflow' model.  
The density (color) and velocity (arrows) distributions on the $y=0$ plane are plotted. 
The elapsed time after protostar formation $t_{\rm ps}$, that after the cloud collapse begins $t$ and the protostellar mass $M_{\rm ps}$ are described in the upper part of each panel.
The white contour corresponds to the boundary between the infalling $v_r<0$ and outflowing $v_r>0$ matter.   
The box size is the same among the panels. 
}
\label{fig:5}
\end{figure*}

\clearpage

%%%%%%
% Fig. 6
%%%%%%
\begin{figure*}
\begin{center}
\includegraphics[width=1.0\columnwidth]{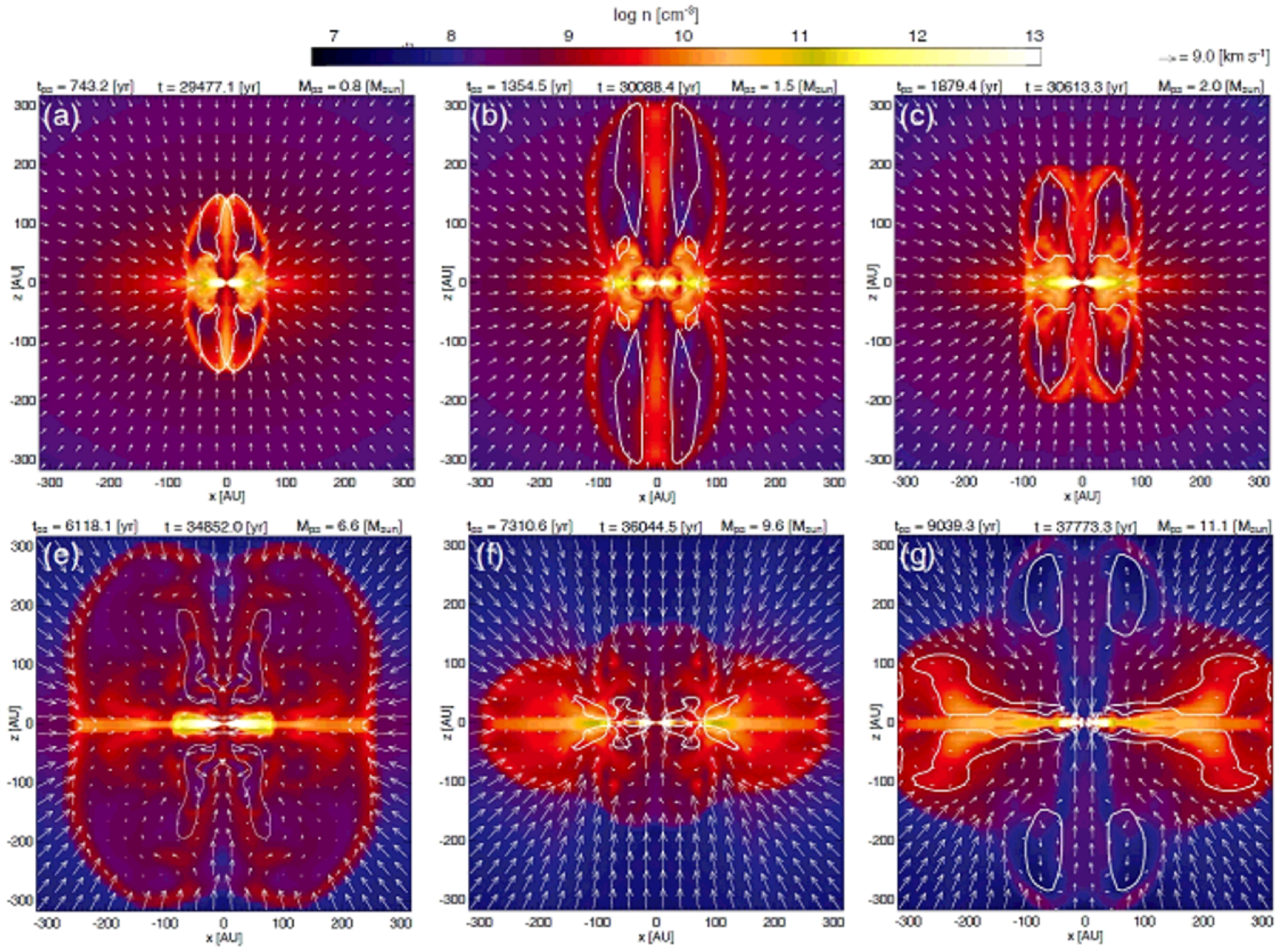}
\end{center}
\caption{
As in Fig.~\ref{fig:5} but for model DM10 ($f=16.8$ and $\mu_0=10$), classified as a `failed outflow' model (see also movie DM10.avi). 
The initial magnetic field strength is set to be half of that for model DM5.
%%, which is highlighted in Fig.~\ref{fig:5}.
The box size is the same among the panels, but differs from that in Fig.~\ref{fig:5}.
}
\label{fig:6}
\end{figure*}
\clearpage

%%%%%%
% Fig. 7
%%%%%%
\begin{figure*}
\begin{center}
\includegraphics[width=1.0\columnwidth]{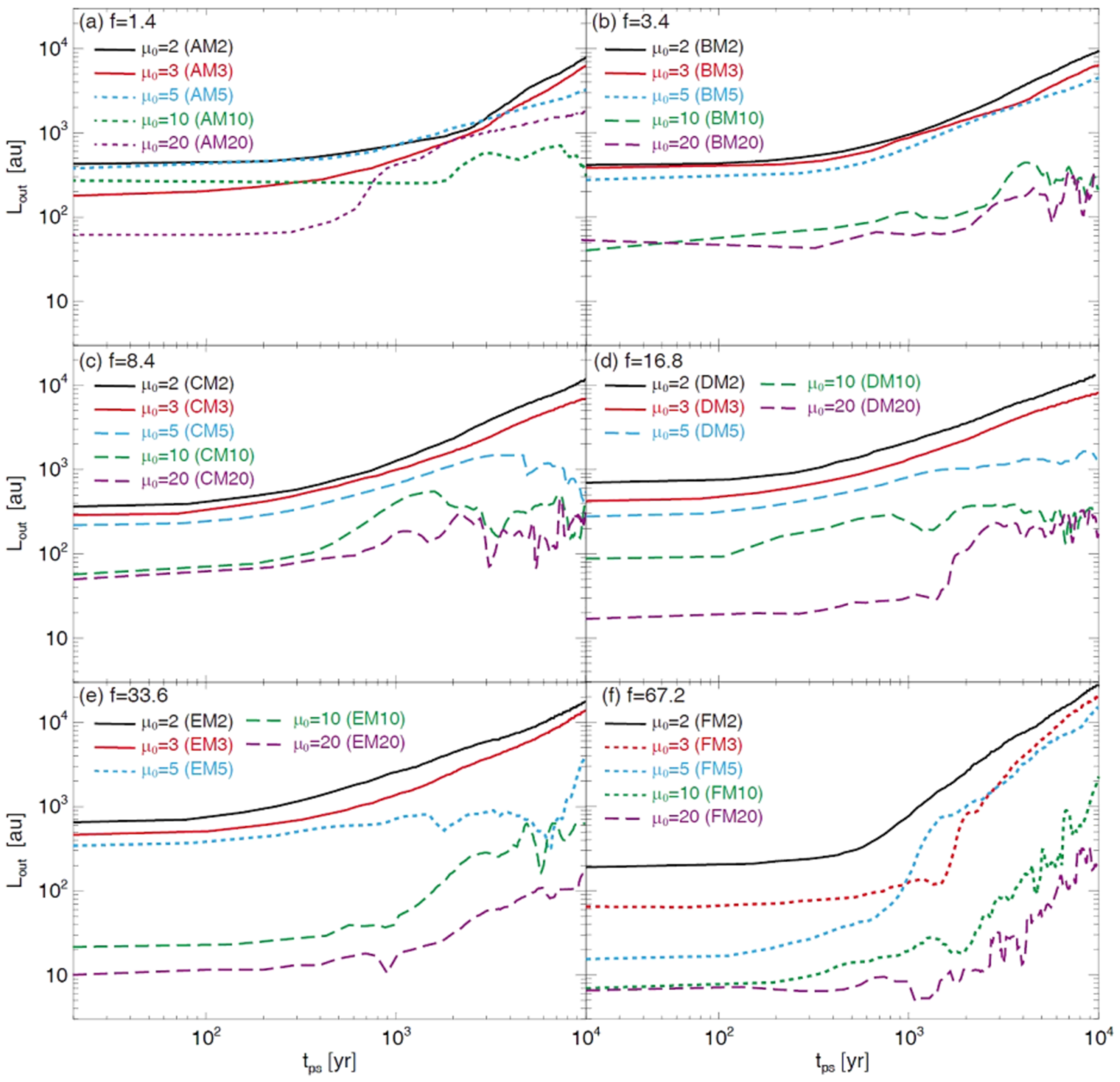}
\end{center}
\caption{
Outflow  size in the vertical direction $L_{\rm out}$ against the elapsed time after protostar formation $t_{\rm ps}$.
In each panel,  models with the same parameter $f$ but different $\mu_0$ are plotted. 
The solid, dotted and broken lines in each panel represent `successful', `delayed' and `failed' outflow models, respectively. 
The model name and parameters $f$ and $\mu_0$ are also described in each panel. 
}
\label{fig:7}
\end{figure*}

%%%%%%
% Fig. 8
%%%%%%
\begin{figure*}
\begin{center}
\includegraphics[width=1.0\columnwidth]{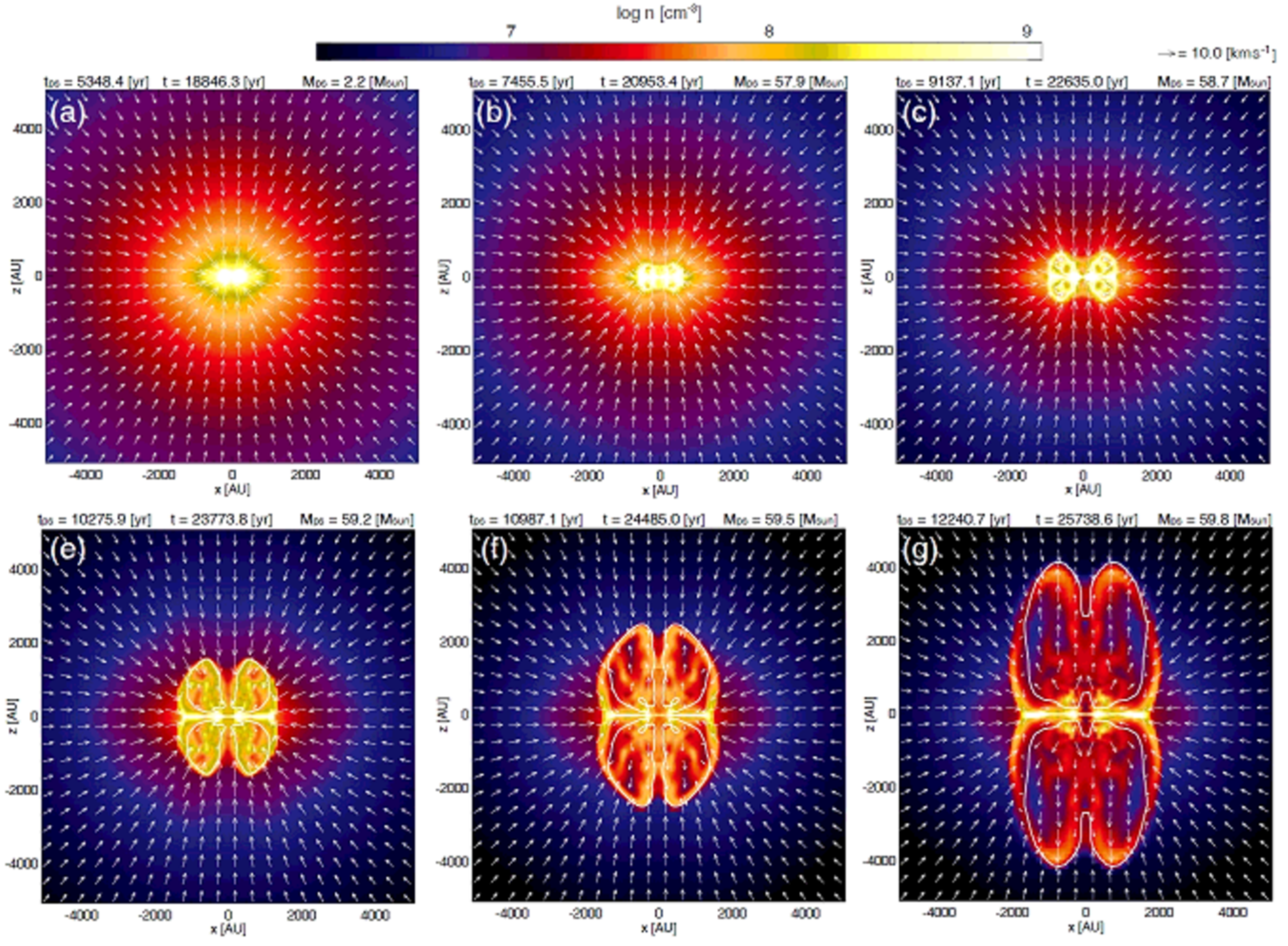}
\end{center}
\caption{
Time sequence of the outflow  for  model FM10 ($f=67.2$ and $\mu_0=10$), classified as a `delayed outflow' model.
The density (color) and velocity (arrows) distributions on the $y=0$ plane are plotted. 
The elapsed time after protostar formation $t_{\rm ps}$, that after the cloud begins to collapse $t$ and the protostellar mass $M_{\rm ps}$ are described above each panel. 
The white contour indicates the boundary between the outflowing and infalling gas inside which the gas is outflowing from the central region.
}
\label{fig:8}
\end{figure*}
\clearpage

%%%%%%
% Fig. 9
%%%%%%
\begin{figure*}
\begin{center}
\includegraphics[width=1.0\columnwidth]{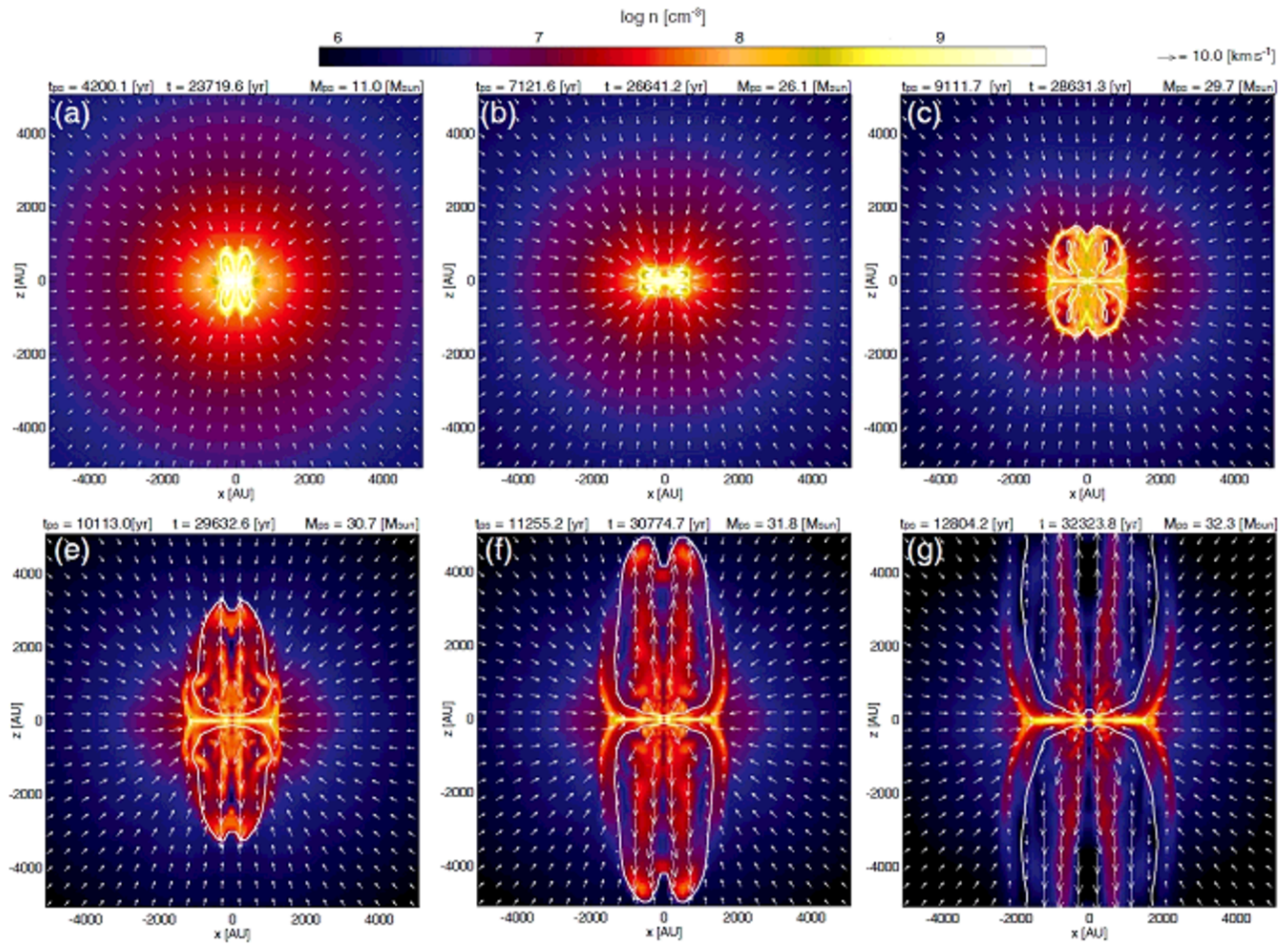}
\end{center}
\caption{
As in Fig.~\ref{fig:8} but for model EM5 ($f=33.6$ and $\mu_0=5$), classified as a `delayed outflow' model (see also movie EM5.avi).
}
\label{fig:9}
\end{figure*}

%%%%%%
% Fig. 10
%%%%%%
\begin{figure*}
\begin{center}
\includegraphics[width=1.0\columnwidth]{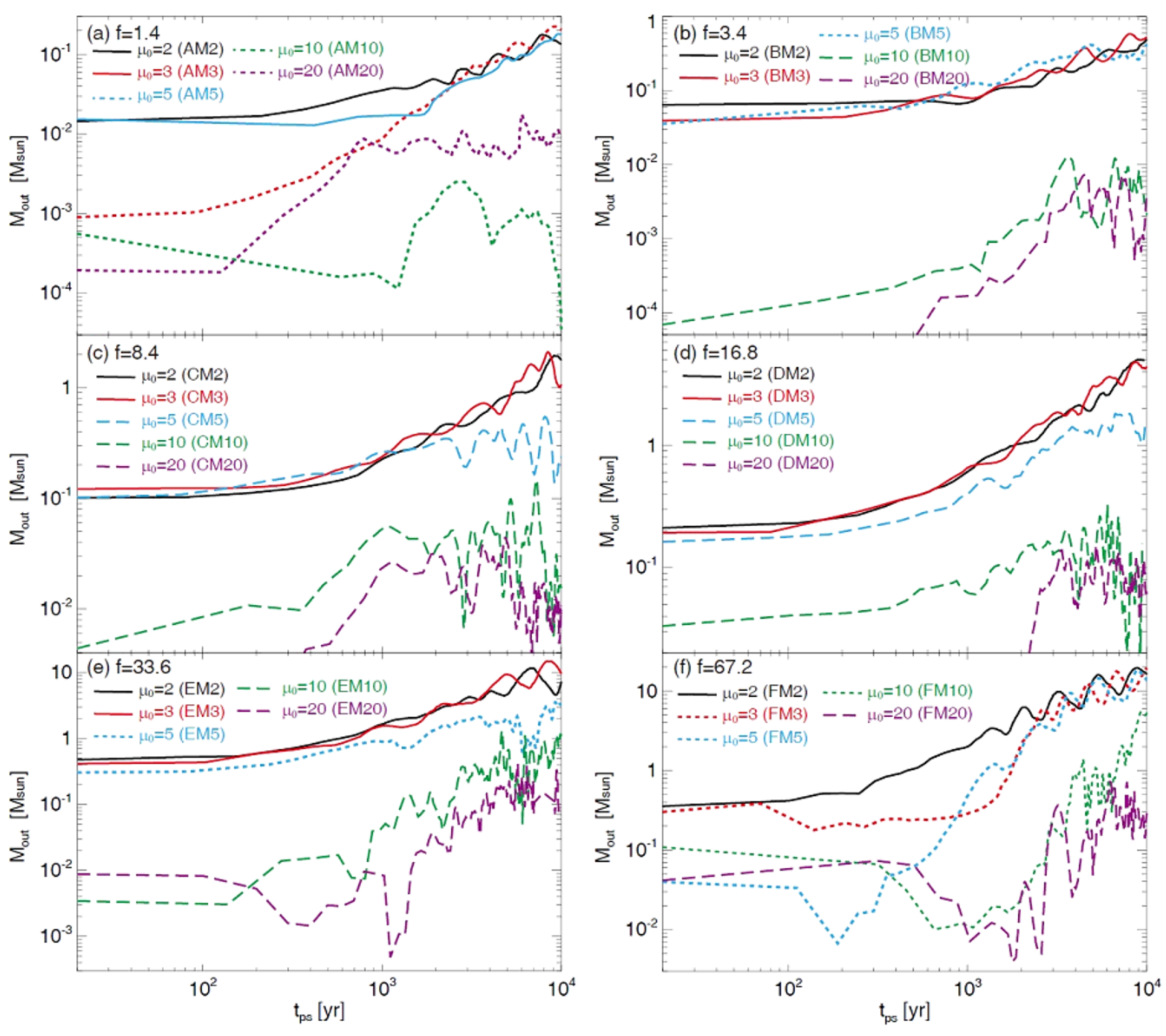}
\end{center}
\caption{
As in Fig.~\ref{fig:7}, but for the outflow mass $M_{\rm out}$.
The solid, dotted and broken lines in each panel represent `successful', `delayed' and `failed' outflow models, respectively.
}
\label{fig:10}
\end{figure*}
\clearpage

%%%%%%
% Fig. 11
%%%%%%
\begin{figure*}
\begin{center}
\includegraphics[width=1.0\columnwidth]{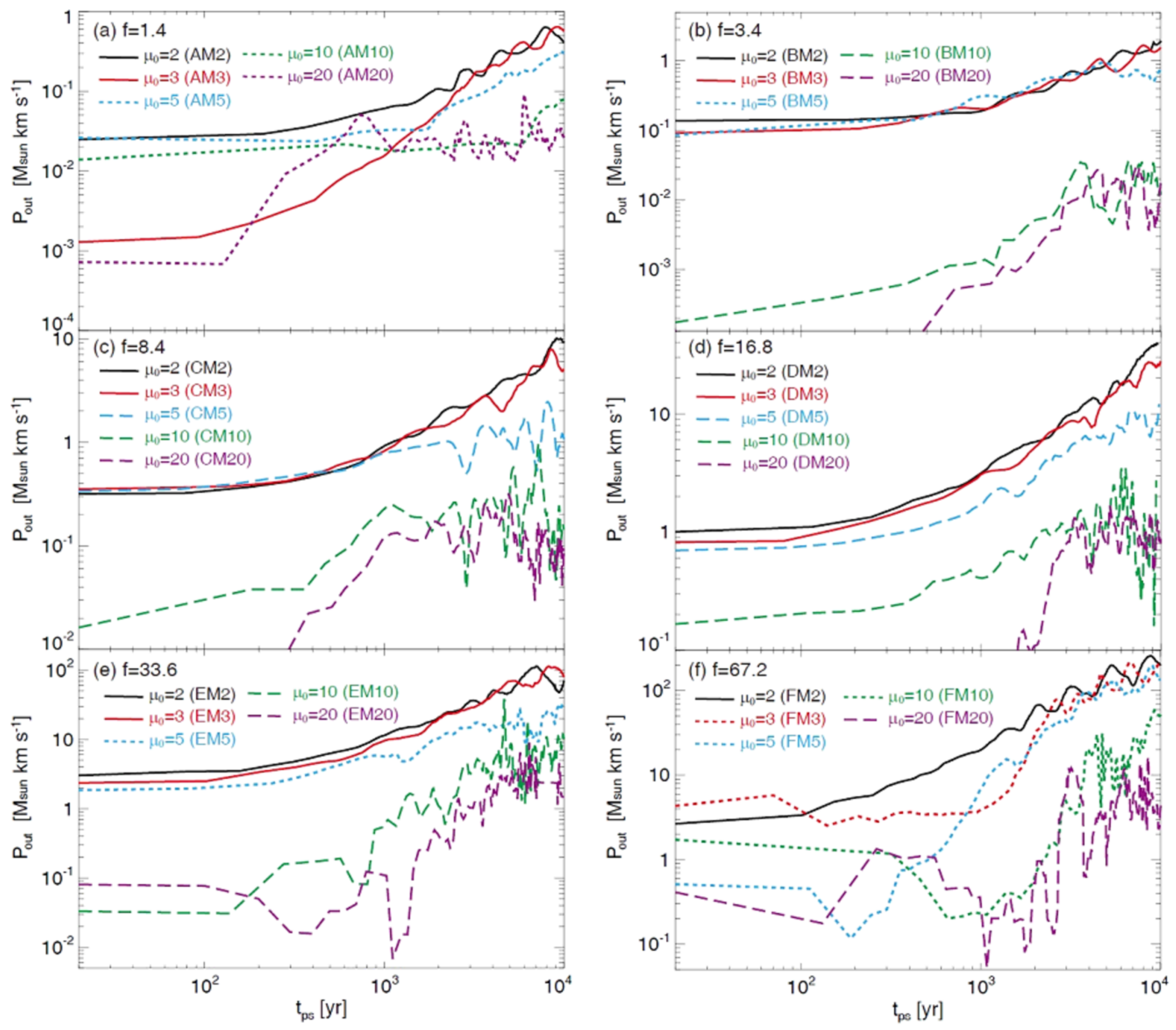}
\end{center}
\caption{
As in Fig.~\ref{fig:7}, but for the outflow momentum $P_{\rm out}$. 
The solid, dotted and broken lines in each panel represent `successful', `delayed' and `failed' outflow models, respectively.
}
\label{fig:11}
\end{figure*}

%%%%%%
% Fig. 12
%%%%%%
\begin{figure*}
\begin{center}
\includegraphics[width=1.0\columnwidth]{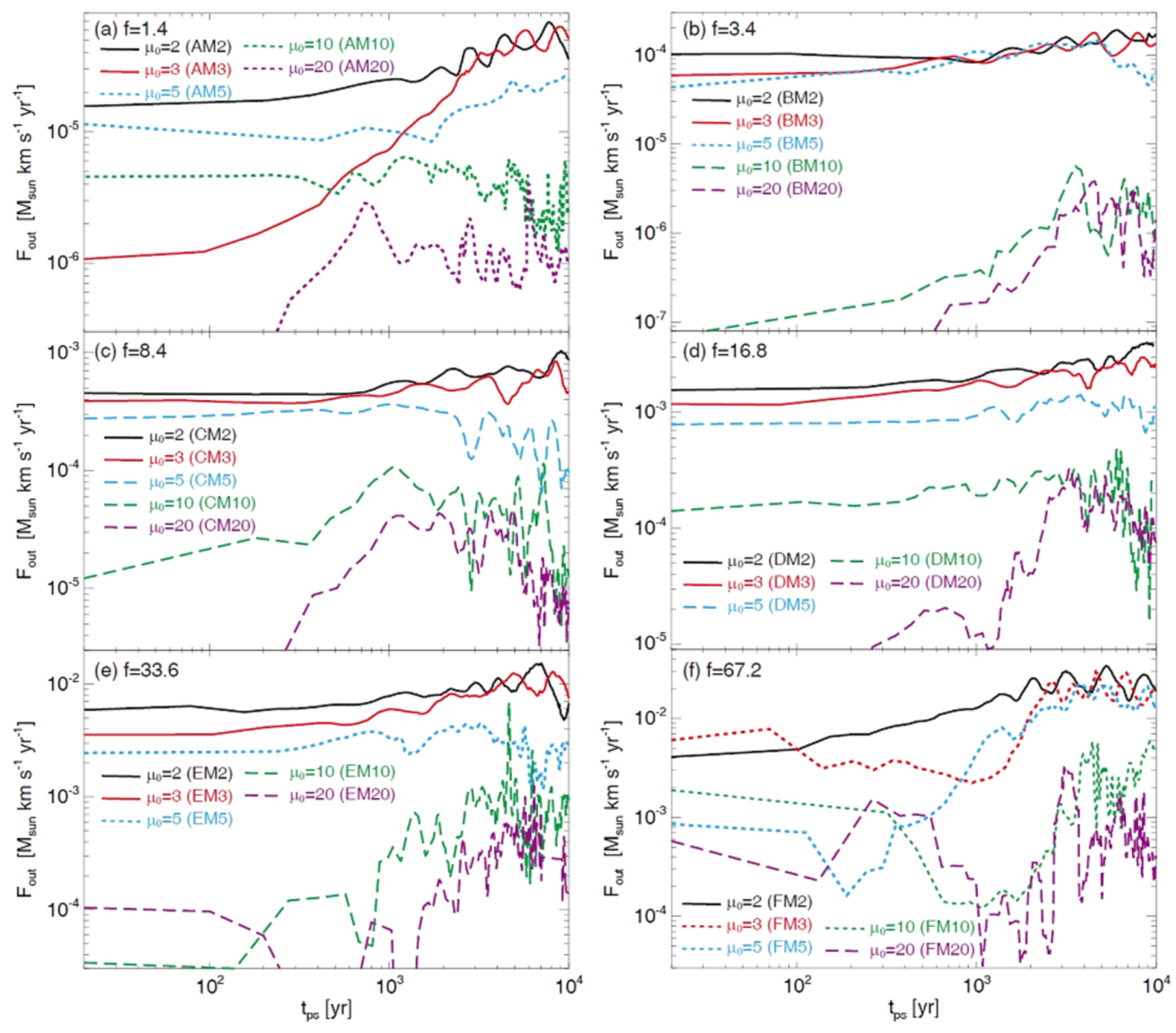}
\end{center}
\caption{
As in Fig.~\ref{fig:7}, but for the outflow momentum Flux $F_{\rm out}$. 
The solid, dotted and broken lines in each panel represent `successful', `delayed' and `failed' outflow models, respectively.
}
\label{fig:12}
\end{figure*}

%%%%%%
% Fig. 13
%%%%%%
\begin{figure*}
\begin{center}
\includegraphics[width=0.95\columnwidth]{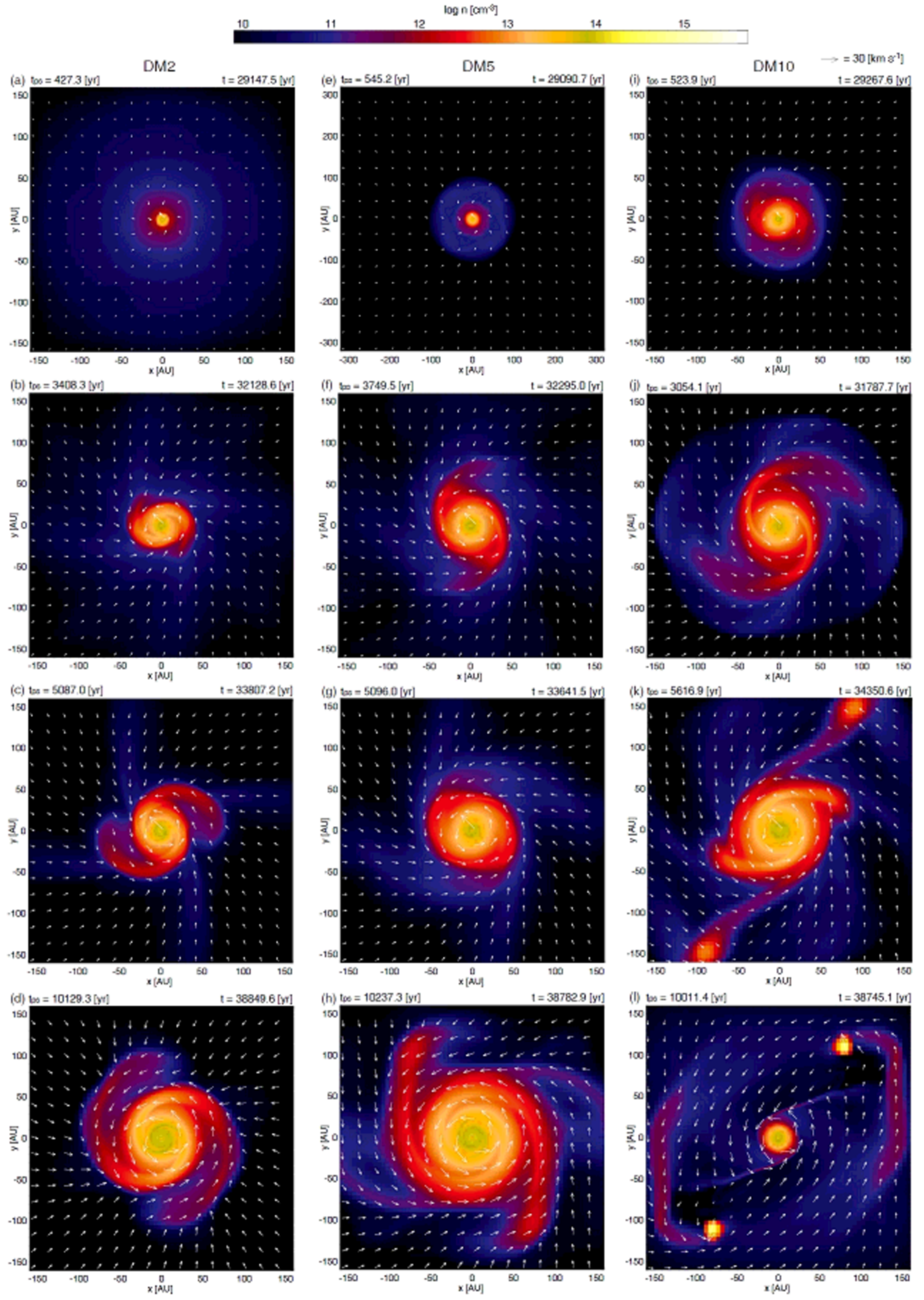}
\end{center}
\caption{
Time sequence of density (color) and velocity (arrows) distributions on the equatorial plane for models DM2 (left; $f=16.8$ and $\mu_0=2$), DM5 (middle; $f=16.8$ and $\mu_0=5$) and DM10 (right; $f=16.8$ and $\mu_0=10$). 
The elapsed time after protostar formation $t_{\rm ps}$ and that after the cloud begins to collapse $t$ are described in each panel. 
The epochs of each model are the same as in Fig.~\ref{fig:4}.
}
\label{fig:13}
\end{figure*}

%%%%%%
% Fig. 13
%%%%%%
%\begin{figure*}
%\begin{center}
%\includegraphics[width=1.0\columnwidth]{f13.eps}
%\end{center}
%\caption{
%Density (color) and velocity (arrows) distributions on the $z=0$ plane for models DM 5 (a), DM10 (b), EM5 (c) and FM10 (d). 
%The first and last two models correspond to `failed outflow' and `delayed outflow' models, respectively.
%The elapsed time after protostar formation $t_{\rm ps}$ and that after the cloud collapse begins $t$ are described in each panel. 
%The protostellar mass $M_{\rm ps}$ is also described in each panel.
%}
%\label{fig:13}
%\end{figure*}

%%%%%%
% Fig. 14
%%%%%%
\begin{figure*}
\begin{center}
\includegraphics[width=1.0\columnwidth]{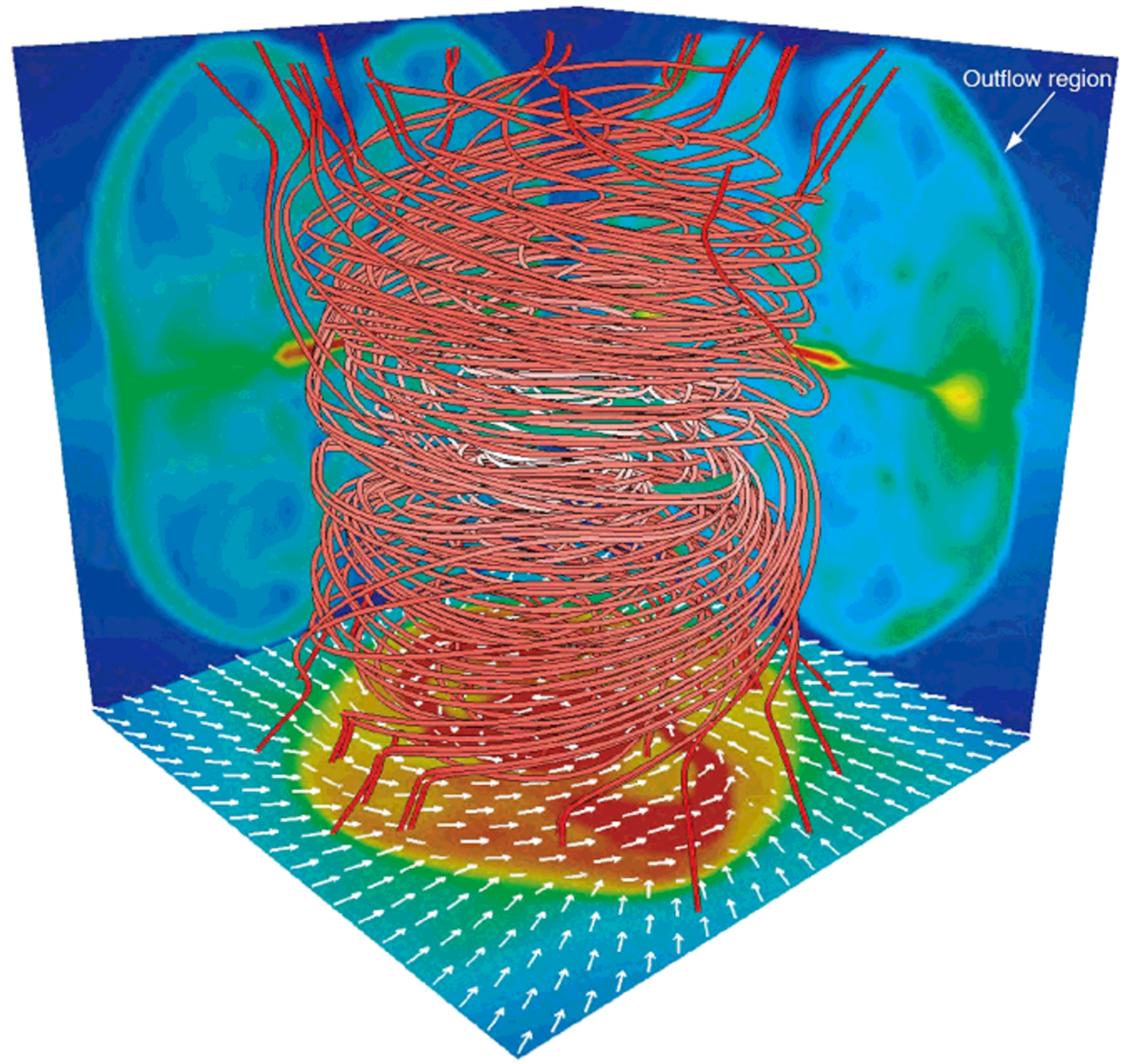}
\end{center}
\caption{
Three dimensional view of the outflow region for model DM10 ($f=16.8$ and $\mu_0=10$) at the same epoch as in Fig.~\ref{fig:6}{\it e} ($t_{\rm ps}=6118.1$\,yr and $t=348520.0$\,yr). 
The red lines correspond to  magnetic field lines. 
The density distributions (color) on the $x=0$, $y=0$ and $z=0$ plane are projected on each wall surface. 
The velocity distribution on the $z=0$ plane is projected on the bottom. 
The outflow region  is indicated by the arrow. 
The box size is 640\,au.
}
\label{fig:14}
\end{figure*}

%%%%%%
% Fig. 15
%%%%%%
\begin{figure*}
\begin{center}
\includegraphics[width=1.0\columnwidth]{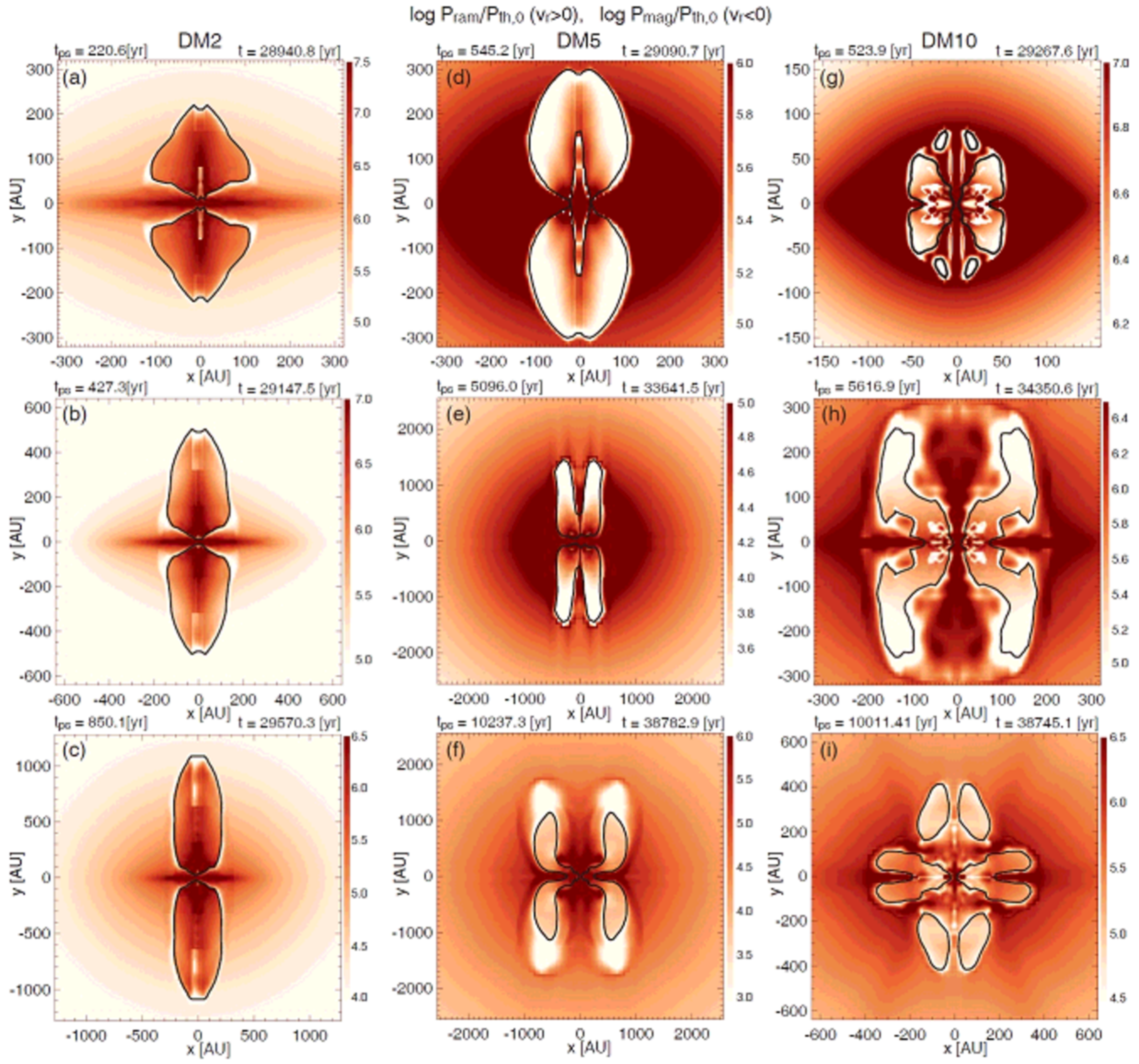}
\end{center}
\caption{
Ram and magnetic pressures on the $y=0$ plane for models DM2 (left), DM5 (middle) and DM10 (right).  
The ram pressure is plotted outside the outflow and the magnetic pressure is plotted inside the outflow. 
Both the ram ($P_{\rm ram}/P_{\rm th, 0}$) and magnetic ($P_{\rm mag}/P_{\rm th, 0}$) pressures are normalized by the initial thermal pressure at the center of the cloud $P_{\rm th, 0}$. 
The boundary between the outflow and infalling envelope is plotted by the black solid curve. 
The elapsed  time after protostar formation $t_{\rm ps}$ and that after the cloud begins to collapse $t$ are described in each panel. 
The spatial scale is different in each panel. 
}
\label{fig:15}
\end{figure*}

%%%%%%
% Fig. 16
%%%%%%
\begin{figure*}
\begin{center}
\includegraphics[width=1.0\columnwidth]{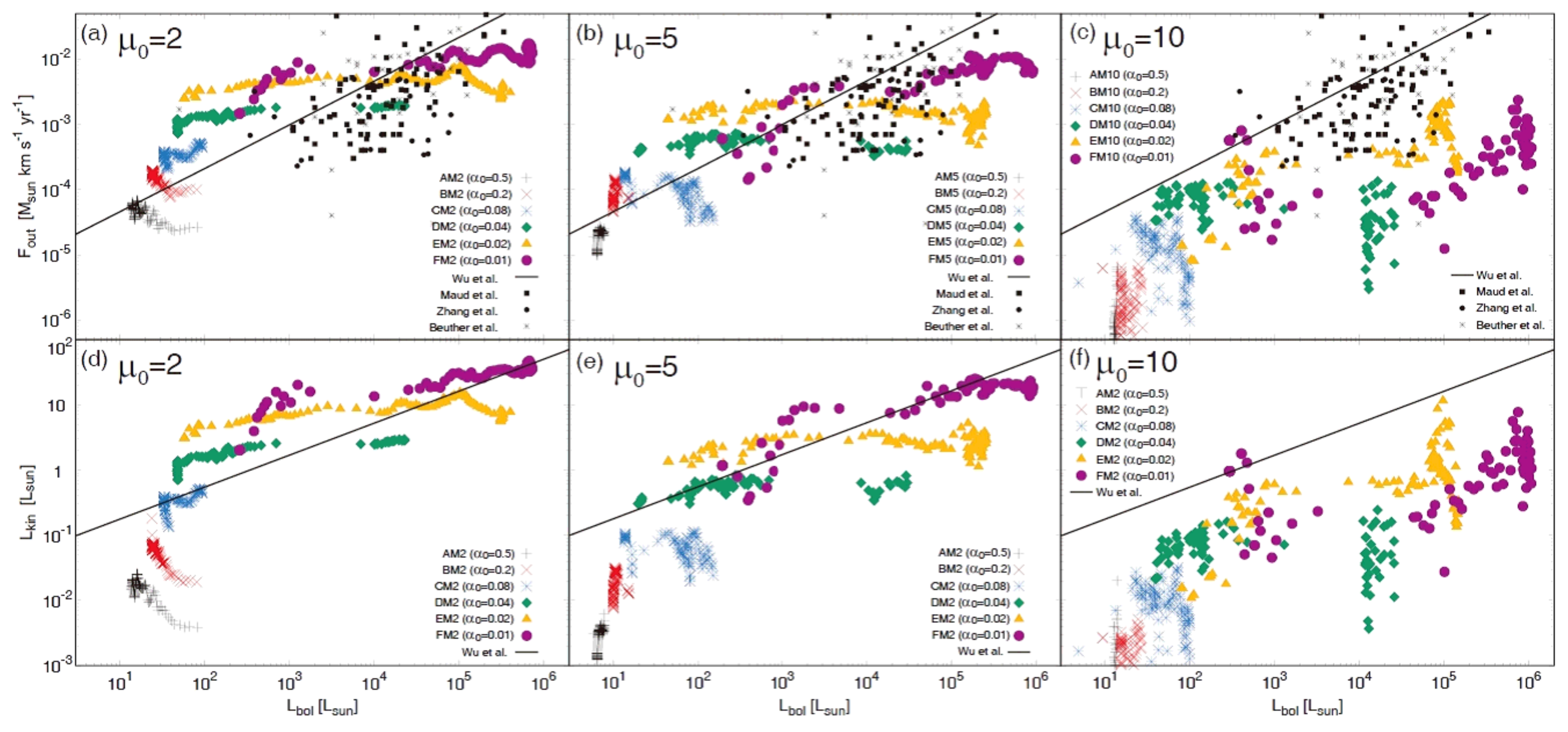}
\end{center}
\caption{
Outflow momentum flux $F_{\rm out}$ (top) and mechanical  luminosity $L_{\rm kin}$ (bottom) against the bolometric luminosity $L_{\rm bol}$ for the models with $\mu_0=2$ (left), 5 (middle) and 10 (right).  
The color points in each panel are the simulation results, for which simulation data are plotted every 100\,yr.
The black solid line in each panel is a fitting formula taken from \citet{wu04}.
The black symbols are taken from observations of  \citet{zhang14}, \citet{maud15} and \citet{beuther02}.
}
\label{fig:16}
\end{figure*}
%%%%%%%%%%%%%%%%%%%%%%%%
\clearpage

\appendix
\section{Protostellar Evolution}
\label{pevol}

\begin{figure*}
\begin{center}
\includegraphics[width=0.5\columnwidth]{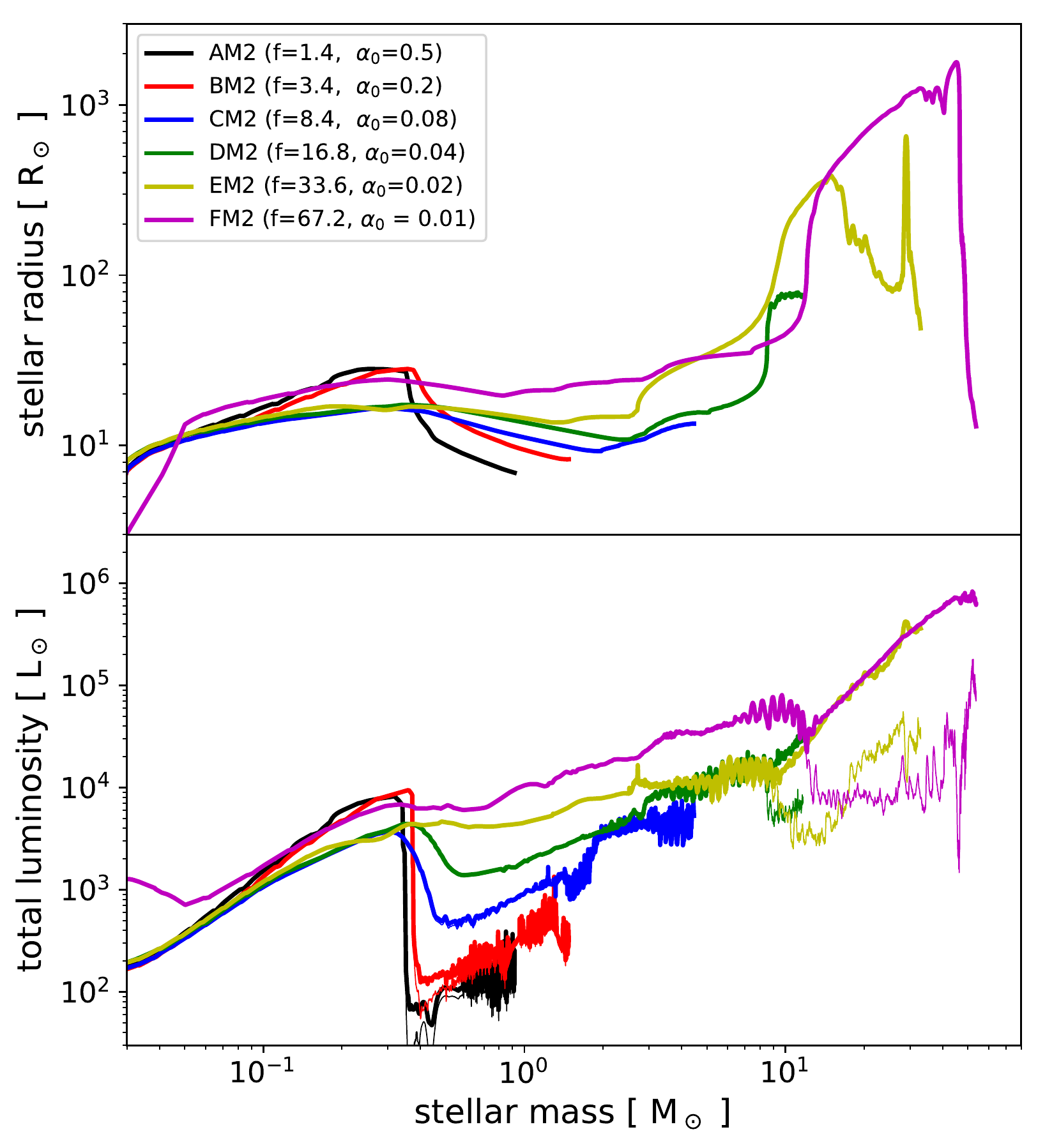}
\end{center}
\caption{
Protostellar evolution calculated under the variable mass accretion histories taken from models AM2, BM2, CM2, DM2, EM2 and FM2, for which the same magnetization parameter $\mu_0 = 2$ is assumed. The top and bottom panels show the evolution of the stellar radius and luminosity as functions of the stellar mass. The different line colors represent the same models as in the upper panels in Fig.~\ref{fig:3}. In the lower panel, the thick and thin lines represent the stellar total and accretion bolometric luminosity, respectively.
}
\label{fig:A1}
\end{figure*}

We briefly describe the protostellar evolution calculated under the variable accretion histories taken from the MHD simulations. Since the accretion histories only weakly depend on the cloud magnetization parameter $\mu_0$ (Fig.~\ref{fig:1}), we only consider the models with $\mu_0 = 2$.
We confirmed that the resultant protostellar evolution is qualitatively the same for the other cases with a different choice of $\mu_0$. Figure~\ref{fig:A1} shows the evolution of the stellar radius and luminosity for cases with different density enhanced factors $f$ (or $\alpha_0$). Since the mean accretion rate is higher for a higher $f$ (Fig.~\ref{fig:1}), the protostellar evolution also varies among the models \citep[e.g.,][]{hosokawa09,hosokawa10}.

The evolution presented here basically follows the trends already found in the previous literature. 
As shown in the upper panel, for instance, the stellar radius tends to be larger for higher accretion rates overall, though it evolves as the stellar mass increases. A striking feature is the rapid swelling that occurs for $M_* \simeq 8$--$15~M_\odot$ in models DM2, EM2 and FM2. In particular, in model FM2, the radius continues to increase even after the swelling, and it eventually exceeds $10^3~R_\odot$. Such an evolution is also known to occur for extremely high accretion rates exceeding $\sim 10^{-2}~M_\odot~{\rm yr}^{-1}$ regardless of the metallicity \citep[e.g.,][]{hosokawa13,haemmerle16}. In model EM2, the star contracts after the swelling and expands again for a short duration at $M_* \simeq 30~M_\odot$. This is caused by an accretion burst event during which the peak rate exceeds $10^{-2}~M_\odot~{\rm yr}^{-1}$.

The evolution of the stellar total luminosity, represented by the sum of the stellar interior luminosity $L_*$ and accretion luminosity $L_{\rm acc}$, also shows characteristic features in the lower panel. The accretion luminosity is the dominant component before the stellar swelling occurs. In models AM2, BM2 and CM2, the bolometric luminosity is well approximated by $L_{\rm acc} \propto M_* \dot{M}_*$ throughout the evolution. The interior luminosity becomes the primary component after the swelling stage, observed in models DM2, EM2 and FM2.

Figure~\ref{fig:A1} also suggests the limitations of our current study. As shown in the upper panel, in models EM2 and FM2, the stellar radius exceeds $\simeq 200~R_\odot \simeq 1$~au for $M_* \gtrsim 10~M_\odot$. Since we adopt the central sink cell whose radius is 1~au (Sec.~\ref{sec:settings}), the actual stellar size exceeds the sink size in the late stage of these models. Such a large stellar radius might affect the outflow launching, which should be further investigated in future simulation studies.

%%%%%%
% Fig. A1
%%%%%%

\section{Disk Properties}
\label{sec:disk}
\begin{figure*}
\begin{center}
\includegraphics[width=1.0\columnwidth]{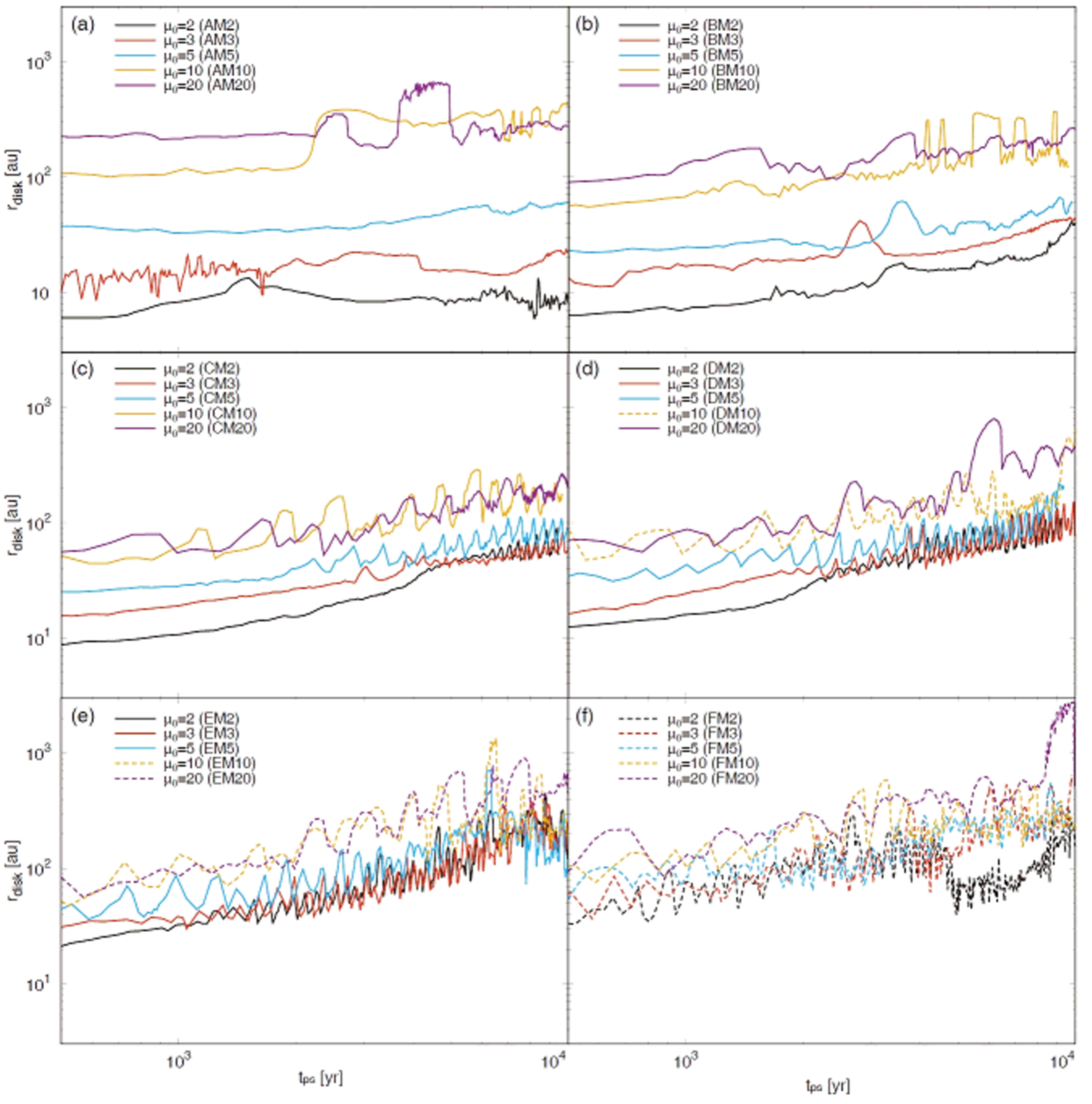}
\end{center}
\caption{
Disk radius $r_{\rm disk}$ against the elapsed time after protostar formation $t_{\rm ps}$.
In each panel,  models with the same parameter $f$ but different $\mu_0$ are plotted. 
The solid and  dotted lines in each panel represent non-fragmentation and fragmentation models, respectively. 
The model name and parameter $\mu_0$ are also described in each panel. 
}
\label{fig:B1}
\end{figure*}

\begin{figure*}
\begin{center}
\includegraphics[width=1.0\columnwidth]{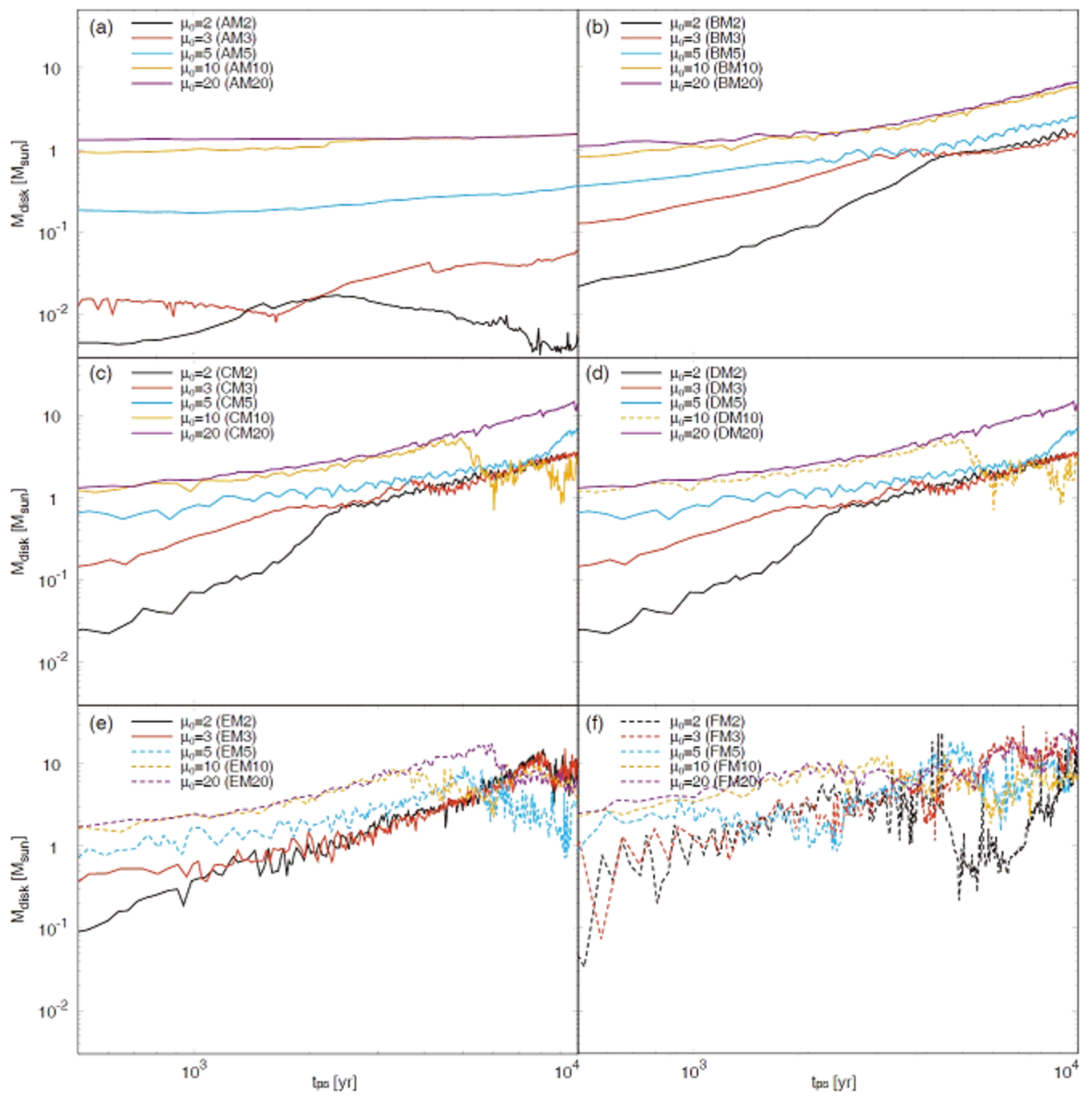}
\end{center}
\caption{
As in Fig.~\ref{fig:B1}, but for the disk mass $M_{\rm disk}$.
The solid and  dotted lines in each panel represent non-fragmentation and fragmentation models, respectively. 
}
\label{fig:B2}
\end{figure*}

In this section, we show the disk radius, mass and Toomre Q parameter of each model to investigate the parameter dependence of the disk properties.
To identify a rotationally supported disk, we adopted  the same prescription used in Paper I (see \S3.3 of Paper I).
The disk radius and mass  for all models are plotted in Figures~\ref{fig:B1} and \ref{fig:B2}, in which the models showing fragmentation are plotted by the dotted line.
The total number of fragments in each model is  described in Table~\ref{fig:1}. 
Note  that the total number means that the number of all fragments appeared in the simulation while many fragments falls onto the center  and a few or no fragment remains at the end of the simulation, as seen in \citet{hosokawa16}.  
When fragmentation occurs and the fragment orbits around the center,  it is difficult to clearly define the disk in our disk identification procedure (see Paper I).
Thus, we do not  focus on the models showing fragmentation to discuss the disk properties.

Figures~\ref{fig:B1} and \ref{fig:B2} show a clear tendency of the disk radius and mass. 
Both the disk radius and mass in the models with initially strong magnetic fields (or smaller  $\mu_0$)  are smaller than those in the models with weak magnetic fields (or larger  $\mu_0$). 
Among  the models with $\alpha_0=0.5$ (models AM2$-$AM20), the disk radius for the strongest magnetic field  model (model AM2, $\mu_0=2$) is about $\sim10$\,au during the simulation, while the disk radii exceed  $\gtrsim 100$\,au for weak magnetic field models AM10 ($\mu_0=10$) and AM20 ($\mu_0=20$).
Thus, the difference in the disk size among these models is over one order of magnitude. 
Although the models with smaller  $\alpha_0$ (or larger mass accretion rate) show the same tendency, the difference in the disk radius is not very significant (Fig.~\ref{fig:B1}{\it c}-{\it f}). 
The disk fragmentation generally occurs with the highest accretion rates (or the smallest $\alpha_0$), regardless of the strength of the magnetic field (Fig.~\ref{fig:B1}{\it f}), as descrbed in Paper I.

We can also confirm the difference in the disk size  in Figure~\ref{fig:13}, in which the disk radius for the models with a strong magnetic  field (left column) is somewhat smaller than that for the models with weak magnetic fields (middle and right columns).
In addition,  a clear spiral structure can been seen in these models (Fig.~\ref{fig:13}). 
Thus, in the models with large accretion rates and weak magnetic fields, it is expected that an excess angular momentum forms a large-sized disk because the angular momentum transfer due to magnetic effects is not effective enough to suppress the disk growth, as described in \S\ref{sec:spiral}. 
Note that a large amount of the mass and angular momentum is introduced into the center in a short duration with a high mass accretion rate (or small $\alpha_0$).
As seen in Figure~\ref{fig:B1}, the disk size is in the range of $\sim 5-1000$\,au for the models without fragmentation.

The disk mass shows the same tendency as the disk radius. 
The disk in the model with a strong magnetic field (or small $\mu_0$) is less massive than that in the model with a weak magnetic field (or large $\mu_0$). 
A sudden drop of the disk mass in Figure~\ref{fig:B2} indicates the epoch when the central star accretes a fragment that reaches the sink after the inward migration.
The small oscillation seen in the radius (Fig.~\ref{fig:B1}) and mass (Fig.~\ref{fig:B2}) is attributed  to the growth of spiral structure, as seen in Figure~\ref{fig:13}. 
After a prominent spiral develops, the disk radius and mass temporarily decrease because the angular momentum is efficiently transferred by the gravitational torque due to the spital structure and the mass accretion is transiently amplified (Fig.~\ref{fig:2}).

\begin{figure*}
\begin{center}
\includegraphics[width=1.0\columnwidth]{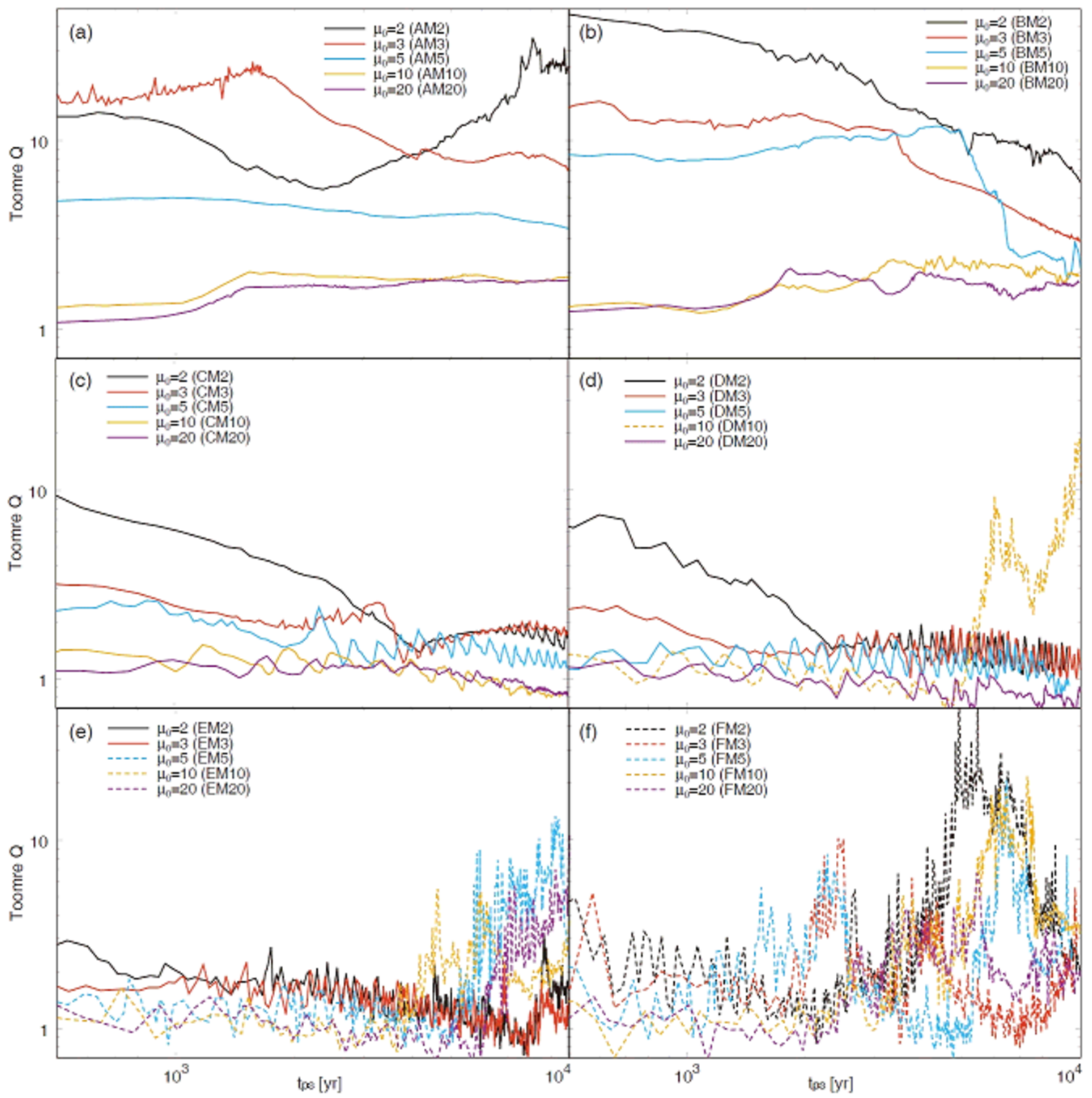}
\end{center}
\caption{
As in Fig.~\ref{fig:B1}, but for  Toomre $Q$ parameter.
The solid and  dotted lines in each panel represent non-fragmentation and fragmentation models, respectively. 
}
\label{fig:B3}
\end{figure*}

\begin{figure*}
\begin{center}
\includegraphics[width=1.0\columnwidth]{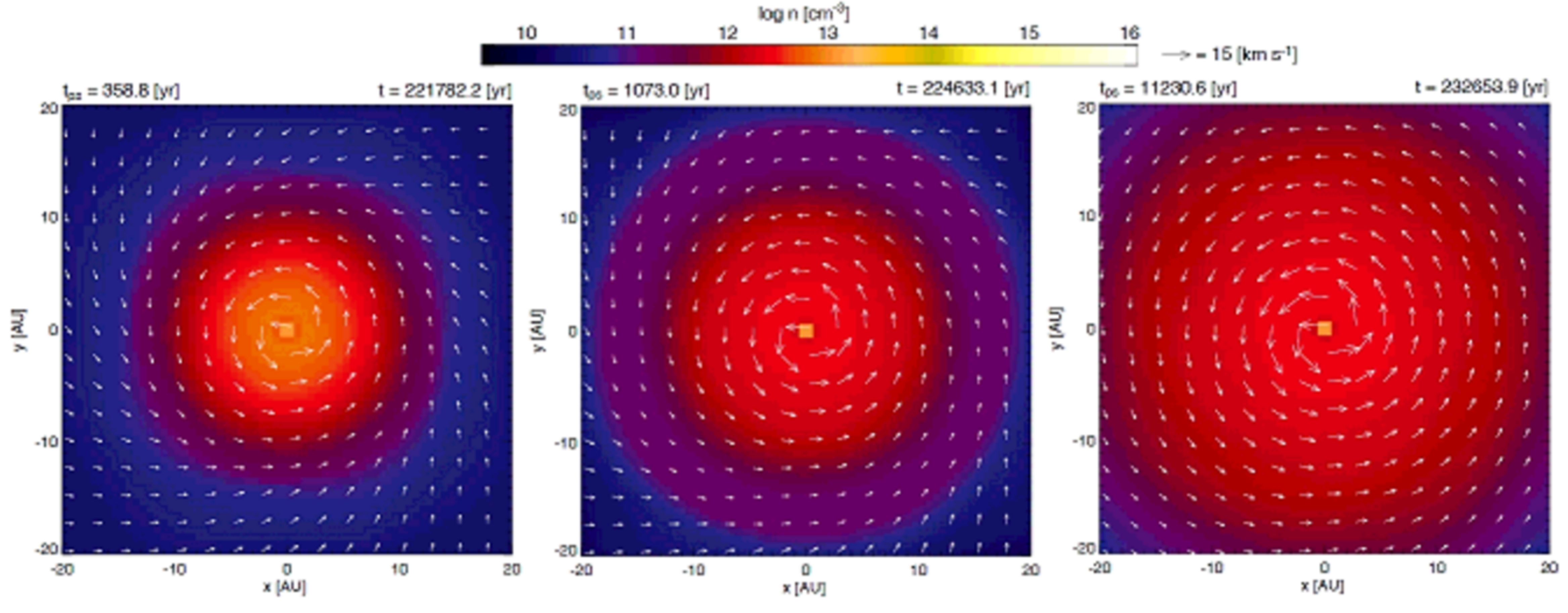}
\end{center}
\caption{
Time sequence of density (color) and velocity (arrows) distributions on the equatorial plane for model AM3 ($f=1.4$ and $\mu_0=3$). 
The elapsed time after protostar formation $t_{\rm ps}$ and that after the cloud begins to collapse $t$ are described in each panel. 
}
\label{fig:B4}
\end{figure*}

To further investigate the gravitational instability, the Toomre $Q$ parameter \citep{toomre64} of each model is plotted in Figure~\ref{fig:B3}. 
Since the derivation of the Toomre $Q$ is the same as in \citet{tomida17}, we simply explain it.
The Q parameter is defined as 
\begin{equation}
< Q >  = \frac{\int_{\rho > \rho_d} \frac{c_s \kappa}{\pi G \Sigma} \Sigma dS}{\int_{\rho > \rho_d} \Sigma dS},
\label{eq:toomre}
\end{equation}
where $c_s$, $\kappa$, $\Sigma$ are the local sound speed, epicyclic frequency and disk surface density, respectively, in which $\kappa = \Omega_{\rm Kep}$ is used assuming the Keplerian disk ($\Omega_{\rm Kep}$ is the Keplerian angular velocity). 
The surface density is derived integrating the density in the vertical direction at each point within the disk (i.e., $\rho > \rho_d$, where $\rho_d$ is the disk critical density, see \citealt{tomida17}).
As seen in equation~(\ref{eq:toomre}), the Toomore Q parameter is averaged over the whole disk.
Figures~\ref{fig:B3}({\it a}) and ({\it b}) indicate that the Q parameter for the models with strong magnetic fields and large $\alpha_0$ (models AM2, AM3, AM4, BM2, BM3) is as large as $Q>2$. 
For these models, the disk radius and mass are small (Figs.~\ref{fig:B1} and \ref{fig:B2}). 
Thus, it is considered that the angular momentum is effectively transported by the magnetic effects and the disk growth and  gravitational instability are suppressed. 
Figure~\ref{fig:B4} shows the time sequence of the density and velocity distributions for model AM3. 
In the figure, we can confirm that no spiral structure develops by the end of the simulation.
Therefore, the contribution of gravitational torque  for the angular momentum transport is expected to be not large.
For the models with large $Q$,  the spiral structure never develops during the simulation.
On the other hand, with small $Q$, the spiral structure develops and fragmentation sometimes occurs. 
The non-axisymmetric (or spiral) structure amplifies the efficiency of the angular momentum transfer due to the gravitational torque.
Thus, when the outflow does not grow, the gravitational torque should play a significant role in the angular momentum transfer.

%%%%%%
% Fig. B1
%%%%%%

%%%%%%
% Fig. B2
%%%%%%
%%%%%%
% Fig. B3
%%%%%%

%%%%%%
% Fig. B4
%%%%%%
\bsp	% typesetting comment
\label{lastpage}
\end{document}